\documentclass[preprint2]{aastex} 
\usepackage{epsfig,latexsym,graphicx,epsf,subfigure,verbatim}
\usepackage{threeparttable,hhline,longtable,rotating}
\usepackage{txfonts}
\shorttitle{Spectroscopy of SN 1999aa}
\shortauthors{Garavini G. et al.}
\begin{document}
\title{Spectroscopic Observations and Analysis of the Peculiar SN~1999aa}
\author{
G.~Garavini\altaffilmark{1},
G.~Folatelli\altaffilmark{1},
A.~Goobar\altaffilmark{1},
S.~Nobili\altaffilmark{1},
G.~Aldering\altaffilmark{2,3},
A.~Amadon\altaffilmark{4},
R.~Amanullah\altaffilmark{1},
P.~Astier\altaffilmark{5},
C.~Balland\altaffilmark{5,6},
G.~Blanc\altaffilmark{2,7},
M.~S.~Burns\altaffilmark{8},
A.~Conley\altaffilmark{2,3,9},
T.~Dahl\'en\altaffilmark{10,11},
S.~E.~Deustua\altaffilmark{2,12},
R.~Ellis\altaffilmark{13},
S.~Fabbro\altaffilmark{14},
X.~Fan\altaffilmark{15},
B.~Frye\altaffilmark{2},
E.~L.~Gates\altaffilmark{16},
R.~Gibbons\altaffilmark{2},
G.~Goldhaber\altaffilmark{2,9},
B.~Goldman\altaffilmark{17},
D.~E.~Groom\altaffilmark{2},
J.~Haissinski\altaffilmark{18},
D.~Hardin\altaffilmark{5},
I.~M.~Hook\altaffilmark{19},
D.~A.~Howell\altaffilmark{2},
D.~Kasen\altaffilmark{2},
S.~Kent\altaffilmark{20},
A.~G.~Kim\altaffilmark{2},
R.~A.~Knop\altaffilmark{21},
B.~C.~Lee\altaffilmark{2},
C.~Lidman\altaffilmark{22},
J.~Mendez\altaffilmark{23,24},
G.~J.~Miller\altaffilmark{25,26},
M.~Moniez\altaffilmark{18},
A.~Mour\~ao\altaffilmark{14},
H.~Newberg\altaffilmark{27},
P.~E.~Nugent\altaffilmark{2},
R.~Pain\altaffilmark{5},
O.~Perdereau\altaffilmark{18},
S.~Perlmutter\altaffilmark{2},
V.~Prasad\altaffilmark{2},
R.~Quimby\altaffilmark{2},
J.~Raux\altaffilmark{5},
N.~Regnault\altaffilmark{2},
J.~Rich\altaffilmark{4},
G.~T.~Richards \altaffilmark{28},
P.~Ruiz-Lapuente\altaffilmark{24},
G.~Sainton\altaffilmark{5},
B.~E.~Schaefer\altaffilmark{29},
K.~Schahmaneche\altaffilmark{5},
E.~Smith\altaffilmark{21},
A.~L.~Spadafora\altaffilmark{2},
V.~Stanishev\altaffilmark{1},
N.~A.~Walton\altaffilmark{30},
L.~Wang\altaffilmark{2},
W.~M.~Wood-Vasey\altaffilmark{2,9},
(THE SUPERNOVA COSMOLOGY PROJECT).
}

\altaffiltext{1}{Department of Physics, Stockholm University, Albanova
University Center, S-106 91 Stockholm, Sweden}
\altaffiltext{2}{E. O. Lawrence Berkeley National Laboratory, 1
Cyclotron Rd., Berkeley, CA 94720, USA }
\altaffiltext{3}{Visiting Astronomer, Cerro Tololo Interamerican
Observatory, National Optical Astronomy Observatory, which is operated
by the Association of Universities for Research in Astronomy, Inc.
(AURA) under cooperative agreement with the National Science
Foundation.}
\altaffiltext{4}{DAPNIA-SPP, CEA Saclay, 91191 Gif-sur-Yvette, France}
\altaffiltext{5}{LPNHE, CNRS-IN2P3, University of Paris VI \& VII,
Paris, France }
\altaffiltext{6}{Universit\'e Paris Sud, IAS-CNRS, B\^atiment 121,
91405 Orsay Cedex, France}
\altaffiltext{7}{Osservatorio Astronomico di Padova, INAF, vicolo
dell'Osservatorio 5, 35122 Padova, Italy}
\altaffiltext{8}{Colorado College, ~14 East Cache La Poudre St.,
Colorado Springs, CO 80903}
\altaffiltext{9}{Department of Physics, University of California
Berkeley, Berkeley, 94720-7300 CA, USA}
\altaffiltext{10}{Stockholm Observatory, Albanova University Center,
S-106 91 Stockholm, Sweden}
\altaffiltext{11}{Space Telescope Science Institute, 3700 San Martin
Drive, Baltimore, MD 21218, USA}
\altaffiltext{12}{American Astronomical Society, 2000 Florida Ave, NW,
Suite 400, Washington, DC, 20009 USA.}
\altaffiltext{13}{California Institute of Technology, E. California
Blvd, Pasadena, CA 91125, USA}
\altaffiltext{14}{CENTRA-Centro M. de Astrof\'{\i}sica and Department
of Physics, IST, Lisbon, Portugal }
\altaffiltext{15}{Steward Observatory, the University of Arizona,
Tucson , AZ 85721}
\altaffiltext{16}{Lick Observatory, P.O. Box 85, Mount Hamilton, CA
95140}
\altaffiltext{17}{Department of Astronomy, New Mexico State University,
Dept. 4500, P.O. Box 30001, Las Cruces, NM 88011}
\altaffiltext{18}{Laboratoire de l'Acc\'elerateur Lin\'eaire,
IN2P3-CNRS, Universit\'e Paris Sud, B.P. 34, 91898 Orsay Cedex, France}
\altaffiltext{19}{Department of Physics, University of Oxford, Nuclear
\& Astrophysics Laboratory, Keble Road, Oxford, OX1 3RH, UK}
\altaffiltext{20}{Fermi National Accelerator Laboratory, P.O. Box 500,
Batavia, IL 60510}
\altaffiltext{21}{Department of Physics and Astronomy, Vanderbilt
University, Nashville, TN 37240, USA}
\altaffiltext{22}{European Southern Observatory, Alonso de Cordova
3107, Vitacura, Casilla 19001, Santiago 19, Chile }
\altaffiltext{23}{Isaac Newton Group, Apartado de Correos 321, 38780
Santa Cruz de La Palma, Islas Canarias, Spain}
\altaffiltext{24}{Department of Astronomy, University of Barcelona,
Barcelona, Spain }
\altaffiltext{25}{Department of Astronomy, San Diego State University,
5500 Campanile Drive, San Diego, CA 92182-1221 }
\altaffiltext{26}{Department of Astronomy, University of Illinois, 1002
West Green Street Urbana, IL 61801 }
\altaffiltext{27}{Rensselaer Polytechnic Institute, Physics Dept.,
SC1C25, Troy NY 12180, U.S.A.}
\altaffiltext{28}{Princeton University Observatory, Peyton Hall,
Princeton, NJ 08544.}
\altaffiltext{29}{Louisiana State University, Department of Physics and
Astronomy,
Baton Rouge, LA, 70803, USA}
\altaffiltext{30}{Institute of Astronomy, Madingley Road, Cambridge CB3
0HA, UK }

\begin{abstract}
We present an extensive new time-series of spectroscopic data of the
peculiar SN~1999aa in NGC 2595. Our data set includes 25 optical
spectra between $-$11 and +58 days with respect to B-band maximum
light, providing an unusually complete time history. The early spectra
resemble those of a SN~1991T-like object but with a relatively strong
Ca H\&K absorption feature. The first clear sign of Si~{\sc
ii}~$\lambda6355$, characteristic of Type Ia supernovae, is found at
day $-$7 and its velocity remains constant up to at least the first
month after B-band maximum light. The transition to normal-looking
spectra is found to occur earlier than in SN~1991T suggesting
SN~1999aa as a possible link between SN~1991T-like and Branch-normal
supernovae.  Comparing the observations with synthetic spectra, doubly
ionized Fe, Si and Ni are identified at early epochs. These are
characteristic of SN~1991T-like objects.  Furthermore, in the day
$-$11 spectrum, evidence is found for an absorption feature which
could be identified as high velocity C~{\sc ii}~$\lambda$6580 or
H$\alpha$. At the same epoch C~{\sc iii}~$\lambda$4648.8 at
photospheric velocity is probably responsible for the absorption
feature at 4500~\AA.  High velocity Ca is found around maximum light
together with Si~{\sc ii} and Fe~{\sc ii} confined in a narrow
velocity window. Implied constraints on supernovae progenitor systems
and explosion hydrodynamical models are briefly discussed.
\end{abstract}
\section{Introduction}
\label{sec_intro}
The observed homogeneity and brightness of Type Ia supernovae (SNe~Ia)
make them excellent tools for distance estimates over extremely large
distances and hence for measurements of cosmological parameters (see
e.g.
\markcite{1998Natur.391...51P,1998ApJ...493L..53G,1998ApJ...507...46S,1998AJ....116.1009R,1999ApJ...517..565P,2003ApJ...598..102K,2003ApJ...594....1T}{Perlmutter} {et~al.} (1998); {Garnavich} {et~al.} (1998); {Schmidt} {et~al.} (1998); {Riess} {et~al.} (1998); {Perlmutter} {et~al.} (1999); {Knop} {et~al.} (2003); {Tonry} {et~al.} (2003)).
However, cosmological results derived from supernovae rely on the
evidence that distant explosions are similar to well-studied nearby
ones and that they can be calibrated with the same techniques,
e.g. the luminosity-light curve-timescale relation
\markcite{1993ApJ...413L.105P}({Phillips} 1993).  The Supernova Cosmology Project (SCP)
coordinated an extensive campaign to study a large number of z $<$ 0.1
SNe~Ia in the spring of 1999 in order to better understand the
intrinsic properties of SNe Ia and thereby improve cosmological
measurements using them
\markcite{2000coex.conf...75A,2000sgrb.conf...47N}({Aldering} 2000; {Nugent}, {Aldering}, \& {The Nearby  Campaign} 2000). The subject of this
work, SN~1999aa, was one of the SCP targets in that campaign.

At this time, several fundamental questions about Type Ia supernova
physics remain. The nature of the progenitor system is still poorly
constrained as are the details of the explosion and thus the origin of
many of the differences observed among Type Ia supernovae.  Normal
SNe~Ia (sometimes called Branch-normal,
\markcite{1983ApJ...270..123B,1993AJ....106.2383B}{Branch} {et~al.}
(1983); {Branch}, {Fisher}, \& {Nugent} (1993)) present early spectra
dominated by intermediate mass elements (IMEs), such as Mg, Si, S, and
Ca, which are replaced by features due to iron-peak ions (such as Fe,
Co, and Ni) as the spectrum evolves with time. However, peculiar
events such as SN~1991T, SN~1997br and SN~2000cx
\markcite{1992ApJ...384L..15F,1992ApJ...397..304J,1992AJ....103.1632P,1992ApJ...387L..33R}({Filippenko}
{et~al.} 1992; {Jeffery} {et~al.} 1992; {Phillips} {et~al.} 1992;
{Ruiz-Lapuente} {et~al.} 1992) show different characteristics. Their
early spectra have weak IME lines and strong doubly ionized iron lines
while spectra after maximum light look almost completely
normal. Moreover, their light curves are generally characterized by a
slow post maximum decline rate, and thus, SN~1991T-like supernovae are
sometimes called peculiar slow decliners. These characteristics have
been regarded as possible signs of different classes of progenitors.

SN~1999aa exhibited spectral characteristics common both to
Branch-normal and peculiar slow decliner SNe~Ia, and thus has been
proposed as a key object which may help in understanding the physical
origin of the observed diversity
\markcite{Branch:2000zm,Branch:2001jf,2001ApJ...546..734L}(Branch 2000; Branch, Baron, \& Jeffery 2001; {Li} {et~al.} 2001b).

Hydrodynamical models predict atmospheric compositions of supernovae in
good agreement with what is found in observed spectra. None of the
currently available models, though, produces an exhaustive description
of the whole spectral time evolution and none is able to reproduce the
complete range of observed differences among SNe~Ia. For extensive
reviews of the theoretical models and observations see
\markcite{1997ARA&A..35..309F,2000ARA&A..38..191H,2000A&ARv..10..179L,Livio:2000cx,Branch:2001jf}{Filippenko} (1997); {Hillebrandt} \& {Niemeyer} (2000); {Leibundgut} (2000); Livio (2000); Branch {et~al.} (2001).

The most widely accepted model for Type Ia supernovae involves the
thermonuclear disruption of a C+O white dwarf star (WD) accreting
material from a companion star
\markcite{1973ApJ...186.1007W,1982ApJ...257..780N,1984ApJS...54..335I,1985cvlm.proc....1P}({Whelan} \& {Iben} 1973; {Nomoto} 1982; {Iben} \& {Tutukov} 1984; {Paczynski} 1985).
From an observational point of view this conclusion is supported by
the amount of energy released, the absence of hydrogen lines (but see
\markcite{2000ApJS..128..615M}{Marietta}, {Burrows}, \&  {Fryxell} (2000)) and the occurrence of SNe~Ia in
elliptical galaxies exclusive of other types.

A more controversial issue is whether SNe~Ia are the result of
explosions at Chandrasekhar mass
\markcite{1969Ap&SS...5..180A,1969Ap&SS...3..464H,1976Ap&SS..39L..37N,1991A&A...245..114K,1994ApJ...423..371W}({Arnett} 1969; {Hansen} \& {Wheeler} 1969; {Nomoto}, {Sugimoto}, \&  {Neo} 1976; {Khokhlov} 1991; {Woosley} \& {Weaver} 1994)
or at sub-Chandrasekhar mass
\markcite{1990ApJ...354L..53L,1991ApJ...370..272L,1994ApJ...423..371W}({Livne} 1990; {Livne} \& {Glasner} 1991; {Woosley} \& {Weaver} 1994). In
the former, thermonuclear burning of carbon occurs in proximity to the
center of the star and the burning front proceeds outward. In the
latter, helium accreting in the external layer of the supernova,
ignites. A detonation then propagates outward through the He layer and
another inward compressing the C+O nucleus that ignites off-center.

While Chandrasekhar mass models have been successful in reproducing
many of the observed characteristics of Branch-normal and
SN~1991T-like supernovae \markcite{1997ApJ...485..812N,1999MNRAS.304...67F}({Nugent} {et~al.} 1997; {Fisher} {et~al.} 1999), sub-Chandrasekhar
models are in good agreement with fainter explosions such as
SN~1991bg-like objects \markcite{1997ApJ...485..812N}({Nugent} {et~al.} 1997).

Pre-maximum spectra can discriminate among these two scenarios since
sub-Chandrasekhar models have external layers dominated by He and Ni
and do not leave any unburned carbon or produce IMEs at expansion
velocities above 14000 km~s$^{-1}$. The identification of C, or strong
Ca or Si lines in pre-maximum spectra would then rule out this
possibility.

Most of the supernovae observed seem to show characteristics in
agreement with  carbon ignition occurring at the center of the WD.
However, this condition can be the result of both a single degenerate
C+O WD accreting hydrogen from a companion or a merging double
degenerate C+O WD, as sometimes suggested for SN~1991T-like supernovae
\markcite{1999MNRAS.304...67F}({Fisher} {et~al.} 1999).

The detection of narrow hydrogen lines in supernova spectra
\markcite{2003Natur.424..651H}({Hamuy} {et~al.} 2003) (see also a discussion in
\markcite{2003ApJ...594L..93L}{Livio} \& {Riess} (2003)) would favor the single degenerate
scenario but a non-detection cannot rule out these models.
A substantial amount of hydrogen can be removed from the companion
star and get mixed into the exploding WD
\markcite{2000ApJS..128..615M}({Marietta} {et~al.} 2000). In this case the detection of H lines is
expected even in low resolution spectra \markcite{2002ApJ...580..374L}({Lentz} {et~al.} 2002)
and would then exclude a double degenerate progenitor system.

While single degenerate models are currently considered the most
promising, other questions remain open. The hydrodynamics of the
explosion and the details of the flame propagation pose many numerical
and conceptual problems currently under investigation,
e.g. computational resolution in 3D simulations, and details of flame
instabilities, for extensive discussion see e.g.
\markcite{Blinnikov:2002gu,2000ARA&A..38..191H}{Hillebrandt} \&
{Niemeyer} (2000) and {Blinnikov} \& {Sorokina} (2002).  Several
models have been proposed to describe the explosion mechanism. Next,
we highlight the possible spectroscopic observables that could help in
constraining the parameters of the models.

Pure one-dimensional (1D) deflagration models (such as W7,
\markcite{1984ApJ...286..644N}{Nomoto}, {Thielemann}, \&  {Yokoi} (1984)) have carbon present down to
v~$>$~14900~km~s$^{-1}$ while 1D delayed-detonation (DD) models
\markcite{1991A&A...245..114K,1992ApJ...393L..55Y,1994ApJ...423..371W}({Khokhlov} 1991; {Yamaoka} {et~al.} 1992; {Woosley} \& {Weaver} 1994)
burn C almost completely up to v~$\sim$~30000~km~s$^{-1}$. Thus, the
identification of carbon lines at low velocity would disfavor
published DD models. Delayed-detonation models can also produce an
unusually high Doppler blue shift of IMEs by tuning the deflagration to
detonation transition density \markcite{2001ApJ...547..402L}({Lentz} {et~al.} 2001).  Thus, IME
lines confined at higher than normal velocities could be more
naturally described by DD models. Lines of stable Fe and Ni at high
velocity in the spectra prior to maximum are also consistent with the
prediction of a deflagration to detonation transition
\markcite{1999ApJS..125..439I}({Iwamoto} {et~al.} 1999).

The complexity of explosion models has advanced to the level where
deflagration in 3D can be explored
\markcite{Khokhlov:2000gj,2003Sci...299...77G}(Khokhlov 2000; {Gamezo}
{et~al.} 2003). These results motivated a few attempts to investigate
the spectral outcome of these models by means of parametrized
synthetic spectra codes
\markcite{2002ApJ...567.1037T,2003ApJ...588L..29B,2003ApJ...593..788K}({Thomas}
{et~al.} 2002; {Baron}, {Lentz}, \& {Hauschildt} 2003; {Kasen}
{et~al.} 2003). In 3D deflagration, due to the highly convoluted
turbulent flame propagation, heavy mixing of freshly synthesized and
unburned material can occur \markcite{2003Sci...299...77G}({Gamezo}
{et~al.} 2003).  Evidence for C and O lines at low velocities, as well
as clumps of burned material (e.g: IME or Fe) in the external layers
(high velocity), would confirm the plausibility of, and the need for,
such a degree of complexity.  The same mechanism that mixes carbon and
oxygen with IME in 3D calculations could probably also work for
hydrogen.

Polarization measurements in SNe~Ia spectra
\markcite{1996AAS...189.4510W,2001ApJ...556..302H,2003ApJ...593..788K,2003ApJ...591.1110W}({Wang}, {Wheeler}, \&  {Hoeflich} 1996; {Howell} {et~al.} 2001; {Kasen} {et~al.} 2003; {Wang} {et~al.} 2003)
have strengthened the conviction that some degree of asymmetry can be
found in supernova atmospheres. In particular, the high velocity (HV)
component of the Ca~{\sc ii} IR triplet found in some supernovae (e.g:
SN~2000cx, SN~2001el) required 3D simulations to be fully reproduced
\markcite{Thomas:2003ip,2003ApJ...593..788K}(Thomas {et~al.} 2003; {Kasen} {et~al.} 2003).

To explore the results of 3D explosion models, full three-dimensional
radiative transfer calculations of supernova spectra would be
required. However, these are not yet available and 1D parametrized
radiative transfer calculations are still in their infancy. Thus,
direct analysis by means of parametrized codes -- both in 1D and in 3D
-- remains the fastest and most versatile way of testing models
predictions and guiding new developments.

In the present work we present a new, comprehensive spectroscopic
timeseries for SN~1999aa.  We use the direct analysis code SYNOW
\markcite{1990sjws.conf..149J,1997ApJ...481L..89F,1999MNRAS.304...67F}({Jeffery} \& {Branch} 1990; {Fisher} {et~al.} 1997, 1999) as
a tool to describe the data and to identify potentially interesting
features. This, together with the measurements of the velocities
inferred from the minima of the spectral features, is used to
investigate the structure of the expanding atmosphere. In particular,
we have looked for evidence of carbon, oxygen, and hydrogen lines in
early spectra as well as the velocity ranges of iron, nickel and
intermediate mass element lines, in order to try to answer some of the
open questions outlined above.

This paper is organized as follows. 
The spectroscopic data of SN 1999aa, and a short description of the
data reduction scheme are presented in section \ref{sec_data}. The
analysis methodology is introduced in section \ref{modeling}. In
section \ref{sec_comparison} our SN~1999aa spectra are compared with
those of spectroscopically peculiar and normal supernovae taken from
the literature. The synthetic spectra for days $-$11, $-$1, +5, +14 and
+40, produced with the highly parametrized SYNOW code, are discussed
in the same section outlining the spectral peculiarities of this
object. Velocities inferred from several spectral features are
analyzed in section \ref{sec_velocities}. A discussion about possible
implications for supernova models is presented in section
\ref{sec_summary} and our conclusions are given in section
\ref{sec_conclusion}.
\section{Data-set and reduction procedure}
\label{sec_data}
\begin{figure*}
 \includegraphics[width=16cm]{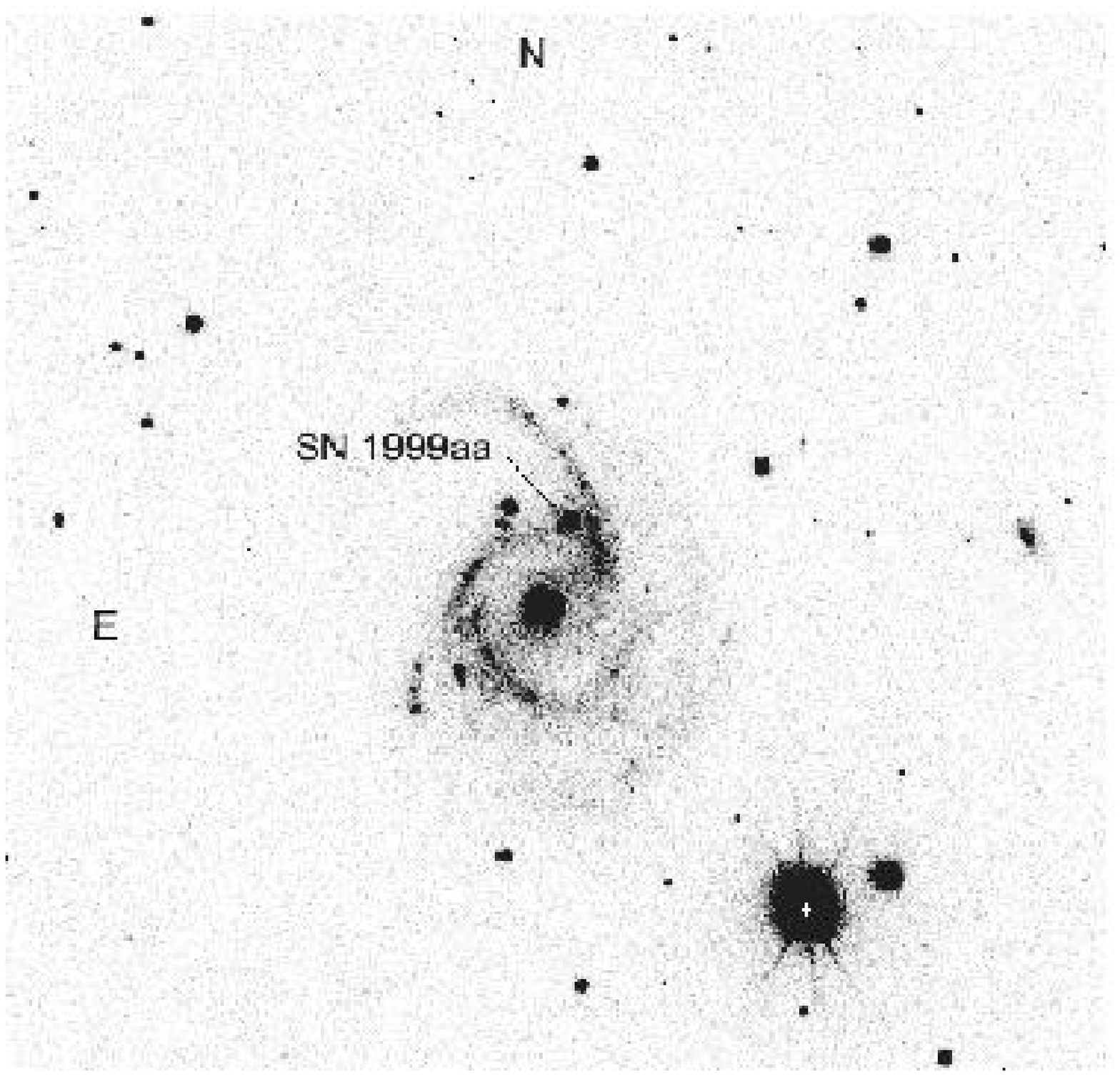}
  \caption{SN~1999aa in its host galaxy NGC 2595.  R.A. =
 $8h27m42\arcsec.03$, Decl. = $+21^{o}29\arcmin14\arcsec.8$ (equinox 2000.0). B-band image obtained at NOT on 1999 February 13 UT with SN~1999aa indicated. The field is 6\arcmin.5 across. North is up, east to the left.}
  \label{99aafinding}
\end{figure*}

 SN1999aa in NGC 2595 (R.A. = $8h27m42\arcsec.03$, Decl. =
 $+21^{o}29\arcmin14\arcsec.8$, equinox J2000.0)  was
 discovered independently by \markcite{1999IAUC.7108....1A}{Armstrong} \& {Schwartz} (1999),
 \markcite{1999IAUC.7109....3Q}{Qiao} {et~al.} (1999) and \markcite{1999IAUC.7109....4N}{Nakano}, {Kushida}, \&  {Kushida} (1999) on
 February 11, 1999. \markcite{1999IAUC.7108....2F}{Filippenko}, {Li}, \&  {Leonard} (1999) noted that it was a
 SN~1991T-like Type Ia supernova based on a spectrum taken the
 following night.  Fig. \ref{99aafinding} shows the position of
 SN~1999aa in its host galaxy NGC~2595, at redshift $z=
 0.0146$. NGC~2595 is an SBc galaxy with a compact blue core and blue
 spiral arms.  The recession velocity of the host galaxy was
 determined from narrow H-alpha and [N ~{\sc ii}] emission
 \markcite{1999IAUC.7108....1A}({Armstrong} \& {Schwartz} 1999).  According to
 \markcite{1998ApJ...500..525S}{Schlegel}, {Finkbeiner}, \&  {Davis} (1998) the Galactic reddening in the direction
 of SN~1999aa is $E(B-V) = 0.04$ mag. Based on the SN colors,
 \markcite{2000ApJ...539..658K}{Krisciunas} {et~al.} (2000) concluded that host galaxy reddening is
 negligible. They also estimated the light curve decline rate $\Delta
 m_{15}$=0.746$\pm$0.024 from a fourth-order polynomial fit to the
 light curve. \markcite{jhathesis}{Jha} (2002) reported $\Delta m_{15}$=0.85$\pm$0.08
 using a different photometry dataset.

The SCP follow-up campaign of this supernova involved 5 different
instruments and resulted in 25 optical spectra ranging between 11 days
before and 58 days after maximum light (all epochs in this work are
given with respect to the B-band maximum).  In most cases the spectra
were acquired using different instrumental settings for the blue and
the red parts of the spectrum in order to avoid possible second order
contamination. Whenever both blue and red spectra were taken at the
same epoch, we present the combined spectrum. The observations were
performed aligning the slit along parallactic angle in order to minimize
light loss from differential atmospheric refraction. Specifications of
the data-set are provided in Table~\ref{tab_data} and the spectral
time sequence is shown in Fig.~\ref{sn99aa.evo.ps}.

All the raw data were analyzed with a common reduction scheme using
standard IRAF routines. The two-dimensional spectra were
bias-subtracted and flat-fielded using calibration images taken with
the same instrument settings as the SN spectra and during the same
night of observation. In most cases, the observations were split into
multiple exposures in order to allow elimination of cosmic rays from
the final spectrum.  Background subtraction was performed on the
resulting spectra using a fitted model of the underlying sky and host
galaxy signal. The supernova spectrum was extracted using the variance
weighted optimal aperture extraction method
\markcite{1986PASP...98..609H}({Horne} 1986). Several arc lamp
exposures, generally taken off supernova position, were used for
wavelength calibration each night.  The accuracy of the calibration
was checked against sky lines and no significant deviations were
found. Atmospheric extinction corrections were applied using tabulated
average extinction curves provided by the observatories.
Spectrophotometric standard stars
\markcite{1990AJ.....99.1621O,1990ApJ...358..344M,1990AJ.....99.1243T,1992PASP..104..533H,1994PASP..106..566H,1995AJ....110.1316B,1996ApJS..107..423S,2001AJ....122.2118B}({Oke}
1990; {Massey} \& {Gronwall} 1990; {Turnshek} {et~al.} 1990; {Hamuy}
{et~al.} 1992, 1994; {Bohlin} {et~al.} 1995; {Stone} 1996; {Bohlin}
{et~al.} 2001) were observed during each night and their latest
calibrations\footnote{We have corrected the
\markcite{1992PASP..104..533H,1994PASP..106..566H}({Hamuy} {et~al.}
1992, 1994) spectra for telluric atmospheric features.} were used to
flux calibrate the SN spectra.  A comparison between observations of
different standard stars during the same night was used to check for
possible systematic errors.  A correction for Galactic extinction was
performed using the standard procedure in
\markcite{1989ApJ...345..245C}{Cardelli}, {Clayton}, \& {Mathis}
(1989) assuming $R_V = 3.1$.  Finally, residual host-galaxy
contamination was checked statistically. The spectrum of the host
galaxy at the position of the supernova, multiplied by an arbitrary
scale factor, was subtracted to the observed data and compared to a
supernova spectral template. The value of the scale factor was
determined by minimizing the $\chi^{2}$ between the model (the
supernova spectral template) and the data (supernova data plus
rescaled host galaxy data).  The host galaxy contamination was found
to be negligible at all studied epochs.  Thus no (additional) galaxy
spectrum was subtracted.  Telluric absorptions and residual fringing
patterns were left uncorrected because they do not affect our present
analysis. For a complete description of the data reduction methodology
see \markcite{method}{Folatelli} (2003).
\section{Spectral Analysis Methodology}
\label{modeling}
As we will use SYNOW to produce synthetic spectra and to investigate
line identifications and velocity ranges of ions in SN1999aa, here we
briefly review its underlying precepts and parameters.  SYNOW
generates spectra using a simple conceptual model of an homologously
expanding supernova envelope.  This model consists of a
continuum-emitting, sharply defined photosphere surrounded by an
extended line-forming, pure scattering atmosphere.  Line transfer is
treated using the Sobolev method
\markcite{1960mes..book.....S,1970MNRAS.149..111C,1990sjws.conf..149J}({Sobolev} 1960; {Castor} 1970; {Jeffery} \& {Branch} 1990).
Thus, line opacity is parametrized in terms of Sobolev optical depth.
The choice of ions used in the calculation is guided by experience and
the SN ion signatures atlas of \markcite{1999ApJS..121..233H}{Hatano} {et~al.} (1999b).  For each
ion introduced, Sobolev optical depth as a function of radius for a
``reference line'' (usually a strong optical line) is specified.
Optical depths in other lines of the ion are set assuming Boltzmann
excitation of the levels at temperature $T_{exc}$.
                                                                                                                                               
 The parameters v$_{phot}$ and $T_{bb}$ set the velocity and blackbody
 continuum temperature of the photosphere, respectively.  For each
 ion, the optical depth $\tau$ and the specified minimum ejection
 velocity v$_{min}$ is given. The optical depth scales exponentially with
 velocity according to $e$-folding velocity, v$_{e}$, up to a maximum
 velocity given by v$_{max}$.  If v$_{min} > $v$_{phot}$ for an ion,
 we refer to the ion as ``detached.''

The black body assumption is a basic simplification of the processes
that contribute to form the continuum emission, therefore, $T_{\rm bb}$
cannot be regarded as physical information.  Thus, SYNOW produces only
a rough indication of the continuum level, which can disagree with the
observed spectrum. We handled this by bringing the blue portions of
the observed and modeled spectra into agreement. The continuum
mismatches which then occur at longer wavelengths have not adversely
affected our study of the few spectral features present in the red.

 Direct analysis results, such as those from SYNOW, yield meaningful
 constraints for hydrodynamical explosion modelers.  Furthermore,
 direct analysis often reveals the presence of lines that otherwise go
 undetected without some treatment of the line blending that
 characterizes supernova spectra.

\section{Spectral Data and Modeling}
\label{sec_comparison}
Prior to B-band maximum light, normal SNe~Ia have spectra dominated by
intermediate mass elements such as Si~{\sc ii}, S~{\sc ii}, Mg~{\sc
ii}, Ca~{\sc ii} and O~{\sc i}
\markcite{1983ApJ...270..123B,1993AJ....106.2383B}({Branch} {et~al.}
1983, 1993). With time, the absorption features from these elements
become weaker and increasingly contaminated by iron-peak element
lines. The substitution is usually complete around 30 days after
maximum, when the photosphere of the supernova starts receding into
the iron-peak core (i.e. the innermost part of the ejecta where mainly
iron-peak elements are present).  Objects such as SN~1991T, SN~1997br
and SN~2000cx represent a deviation from the impressive homogeneity of
the spectral and photometric characteristics of SN explosions
\markcite{1999MNRAS.304...67F,1999AJ....117.2709L,2001PASP..113.1178L}({Fisher}
{et~al.} 1999; {Li} {et~al.} 1999, 2001a). This has raised the
question whether they could be explained as a different physical
phenomenon or as extreme cases of the same process. Ever since the
first spectra of SN~1999aa started circulating in the supernova
community, several authors suggested that this object could be helpful
in addressing this issue
\markcite{2001PASP..113..169B,2001ApJ...546..734L}({Branch} 2001; {Li}
{et~al.} 2001b).  This work should be viewed in that context.

In order to understand and interpret the differences between SN~1999aa
and other normal and peculiar supernovae we now analyze the spectral
time evolution through spectral comparison and spectral modeling. We
have selected five epochs, approximately one week apart from each
other, to describe the spectral time evolution in all interesting
phases.  These are: $-$11, $-$1, +5, +14, and +40 days with respect to
B-band maximum.

\subsection{Day 11 before maximum}
\begin{figure*}
\centering 
 \includegraphics[width=16cm]{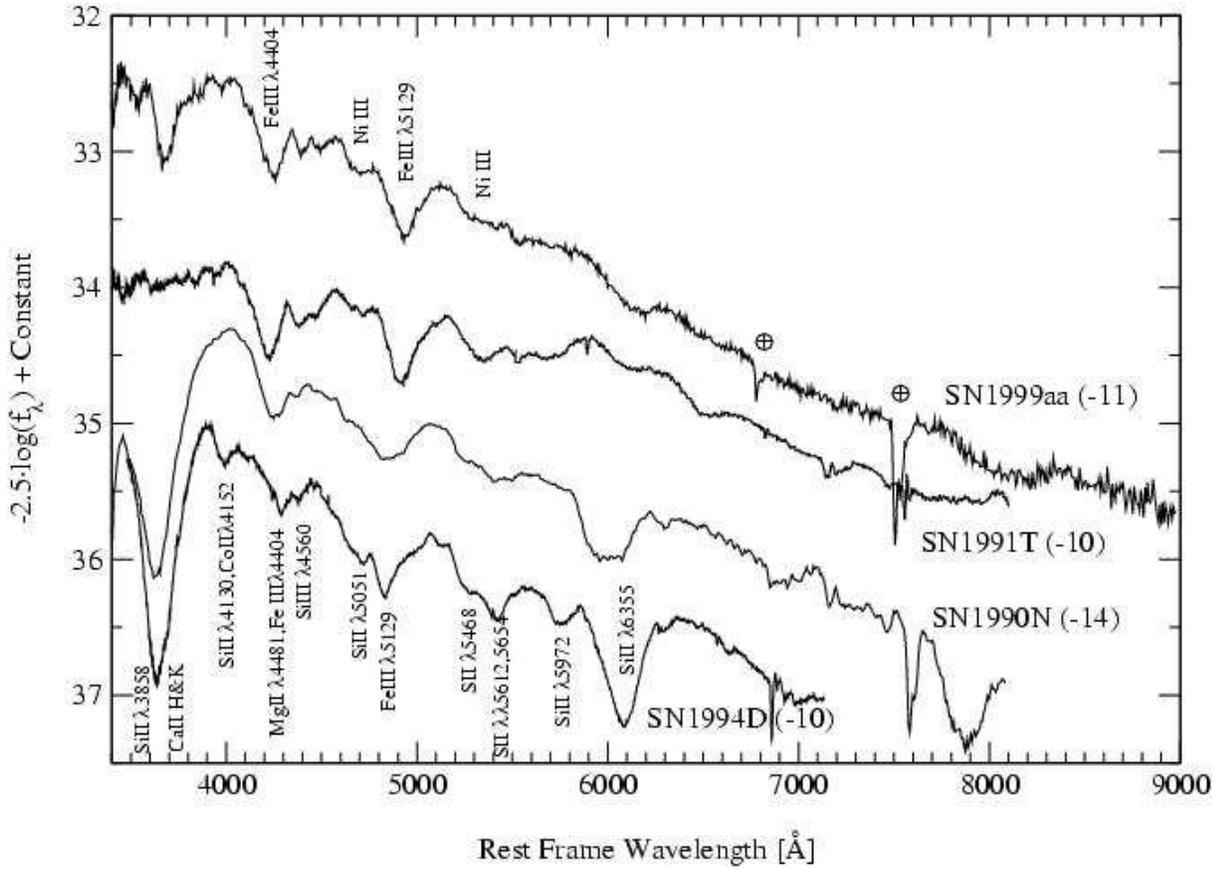}
  \caption{The day $-$11 spectrum of SN~1999aa together with those of
   SN~1991T, and normal SN~1990N and SN~1994D from
   \markcite{1992ApJ...384L..15F,1991ApJ...371L..23L,1996MNRAS.278..111P}{Filippenko}
   {et~al.} (1992); {Leibundgut} {et~al.} (1991); {Patat} {et~al.}
   (1996). Each spectrum is labeled with the phase (days since B
   maximum). Line identification is explained in the text. The
   $\oplus$ symbol marks atmospheric absorptions.}
  \label{comp-11.ps}
\end{figure*}

Our spectrum of SN~1999aa at 11 days prior to B-band maximum light is
shown in Fig.~\ref{comp-11.ps}. The spectra of SN~1991T, SN~1990N and
SN~1994D are also shown for comparison. The identification of the
lines labeled here and in the following graphs follow those in
\markcite{1999AJ....117.2709L,2001PASP..113.1178L,1999MNRAS.304...67F,1996MNRAS.278..111P,1995A&A...297..509M,1993ApJ...415..589K,1992ApJ...397..304J}{Li}
{et~al.} (1999, 2001a); {Fisher} {et~al.} (1999); {Patat} {et~al.}
(1996); {Mazzali}, {Danziger}, \& {Turatto} (1995); {Kirshner}
{et~al.} (1993); {Jeffery} {et~al.} (1992). The spectra of SN~1999aa
and SN~1991T are very different from those of normal supernovae
SN~1990N and SN~1994D around 11 days prior to the B-band maximum
light.  Instead of the typical Si~{\sc ii}, S~{\sc ii} and Mg~{\sc ii}
lines, early spectra of these peculiar SNe~Ia are dominated by two
deep absorptions due to Fe~{\sc iii}~$\lambda\lambda$4404 and
5129. Si~{\sc iii}~$\lambda$4560 is responsible for the smooth
absorption on the red side of the Fe~{\sc iii}~$\lambda4404$ line.
Possibly weaker Ni~{\sc iii} lines contribute near 4700~\AA\ and
5300~\AA, \markcite{1992ApJ...397..304J,1992ApJ...387L..33R}({Jeffery}
{et~al.} 1992; {Ruiz-Lapuente} {et~al.} 1992).  What mainly
distinguishes SN~1999aa from SN~1991T is the presence in the former of
the trough around 3800~\AA, most probably due to Ca~{\sc ii}
H\&K. This absorption is weak compared to normal SNe~Ia. The weak and
broad line around 6150~\AA\ could be due to Si~{\sc ii}~$\lambda$6355
with some contamination from C~{\sc ii}~$\lambda$6580.

\subsection*{Synthetic spectrum}
\label{synow-11_s}
\begin{figure*}
\centering
 \includegraphics[width=16cm]{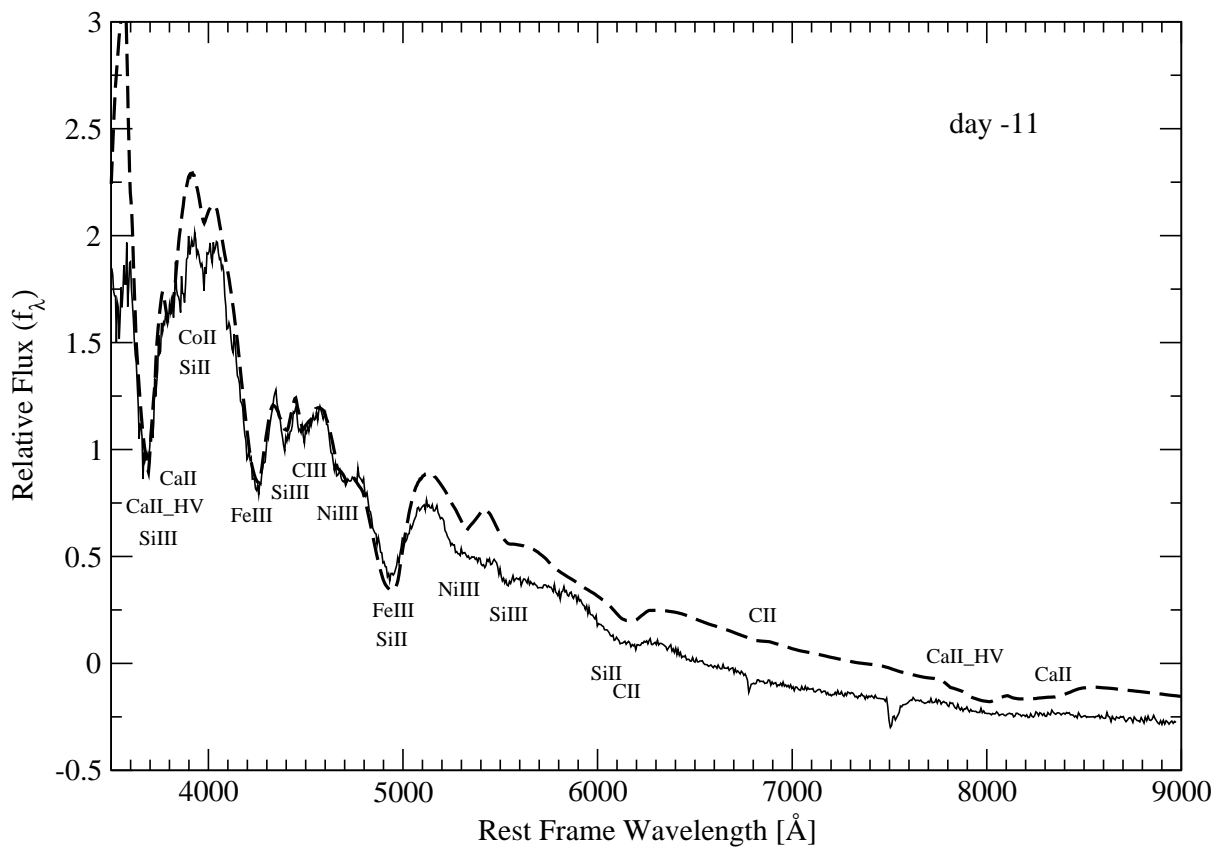}
  \caption{Synthetic spectrum compared with SN~1999aa spectrum for day
  $-$11. SYNOW parameters are presented in Table \ref{table-11}. The
  region around 6150~\AA\ and 4500~\AA\ are enlarged in
  Figs. \ref{synow-11CII} and \ref{synow-11CIII} to highlight the
  possible contributions of C~{\sc ii} and C~{\sc iii}. The mismatch
  of the continuum level above 5000~\AA\ has to be considered a
  limitation of the underlying black body assumption but does not
  affect the line identification. Ions responsible for features in the
  synthetic spectrum are labeled.}
  \label{synow-11}
\end{figure*}

In Fig.~\ref{synow-11} we show the results for the best match
synthetic spectrum computed using the parameters in Table
\ref{table-11}.  The continuum black body temperature is set to
13700~K and the photosphere is placed at 11000~km~s$^{-1}$.  The
dominant and characteristic features at this epoch are due to Fe~{\sc
iii} lines.  While some authors identify the small absorption line at
4000~\AA\ only as Si~{\sc ii}~$\lambda$4130, in our synthetic spectrum
we find that Si~{\sc ii} alone cannot reproduce the feature completely
due to its small optical depth.  If we add a contribution from Co~{\sc
ii} (as in \markcite{1999ApJ...525..881H}{Hatano} {et~al.} (1999a) for example) we are able to
reproduce the observed spectrum. Note that to improve the match
Co~{\sc ii} needs to be detached above the photosphere. Si~{\sc iii}
appears confined below $16\times 10^{3}$ km~s$^{-1}$ (similarly to
Si~{\sc ii}) and is responsible for the features at $~4400$~\AA, and
$~5500$~\AA, as originally identified by \markcite{1992ApJ...397..304J}{Jeffery} {et~al.} (1992)
for SN~1991T and SN~1990N.

The only way we could account for the complete structure of the
Ca~{\sc ii} H\&K features and for its velocity distribution (blue edge
at 40000~km~s$^{-1}$) was to introduce a HV component (with
higher ${\rm v}_{\rm e}$) in addition to the one at ${\rm v}_{\rm
max}$=19500~km~s$^{-1}$. This allows us to `fill-up' the whole
spectral profile at this epoch and at day $-$1 it will produce the
observed blue absorption feature of the Ca~{\sc ii} IR triplet, as
noticed already in \markcite{1999ApJ...525..881H}{Hatano} {et~al.} (1999a) for SN~1994D and
analyzed in \markcite{2003ApJ...593..788K}{Kasen} {et~al.} (2003) for SN~2001el and in
\markcite{2003astro.ph..2260T}{Thomas} {et~al.} (2003) for SN~2000cx. The region around 4600~\AA\
and the broad absorption at 5300~\AA\ can be matched with detached
Ni~{\sc iii}.

We have focused in on the wavelength regions around 6150~\AA\ and
4500~\AA\ for particular study, since these have been discussed for
other SNe~Ia in the literature.  Si~{\sc ii} $6150$ has a very broad
profile. As we will show in section \ref{sec_velocities} the velocity
fitted for Si~{\sc ii}~$6355$ at this epoch is the lowest among all
known supernovae (v$\sim$8000 km s$^{-1}$). This suggests that
probably another ion is partially or totally responsible for this
absorption feature. In Fig.~\ref{comp-11.ps}, a similar absorption
feature is visible for the spectrum at $-$10 days of
SN~1991T. \markcite{1999MNRAS.304...67F}{Fisher} {et~al.} (1999) discussed the possible presence
of HV C in early spectra of SN~1991T based on the line at
6150~\AA\ usually attributed to Si~{\sc ii} $6355$.  The same
possibility has been also proposed for SN~1990N in
\markcite{1997ApJ...481L..89F}{Fisher} {et~al.} (1997) and \markcite{2001MNRAS.321..341M}{Mazzali} (2001), for
SN~1994D in \markcite{1999ApJ...525..881H}{Hatano} {et~al.} (1999a) and recently for SN~1998aq in
\markcite{2003AJ....126.1489B}{Branch} {et~al.} (2003).

\begin{figure*}
\centering  
\includegraphics[width=16cm]{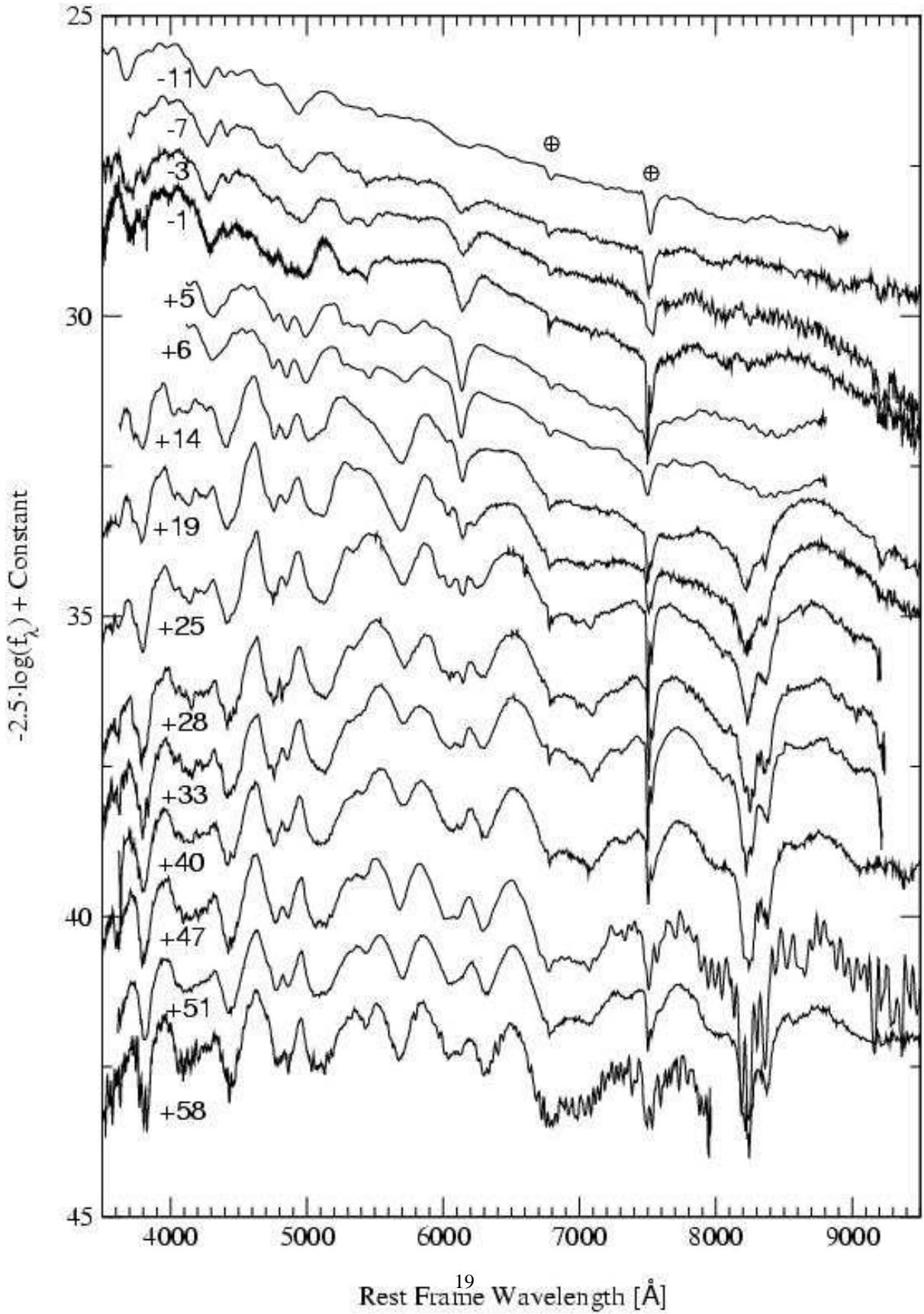}
\caption{SN~1999aa spectral time sequence. The spectra are labeled in
  days relative to B-band maximum light. The $\oplus$ symbol marks
  telluric atmospheric absorptions.}
  \label{sn99aa.evo.ps}
\end{figure*}

\begin{figure*}
\centering
 \includegraphics[width=16cm]{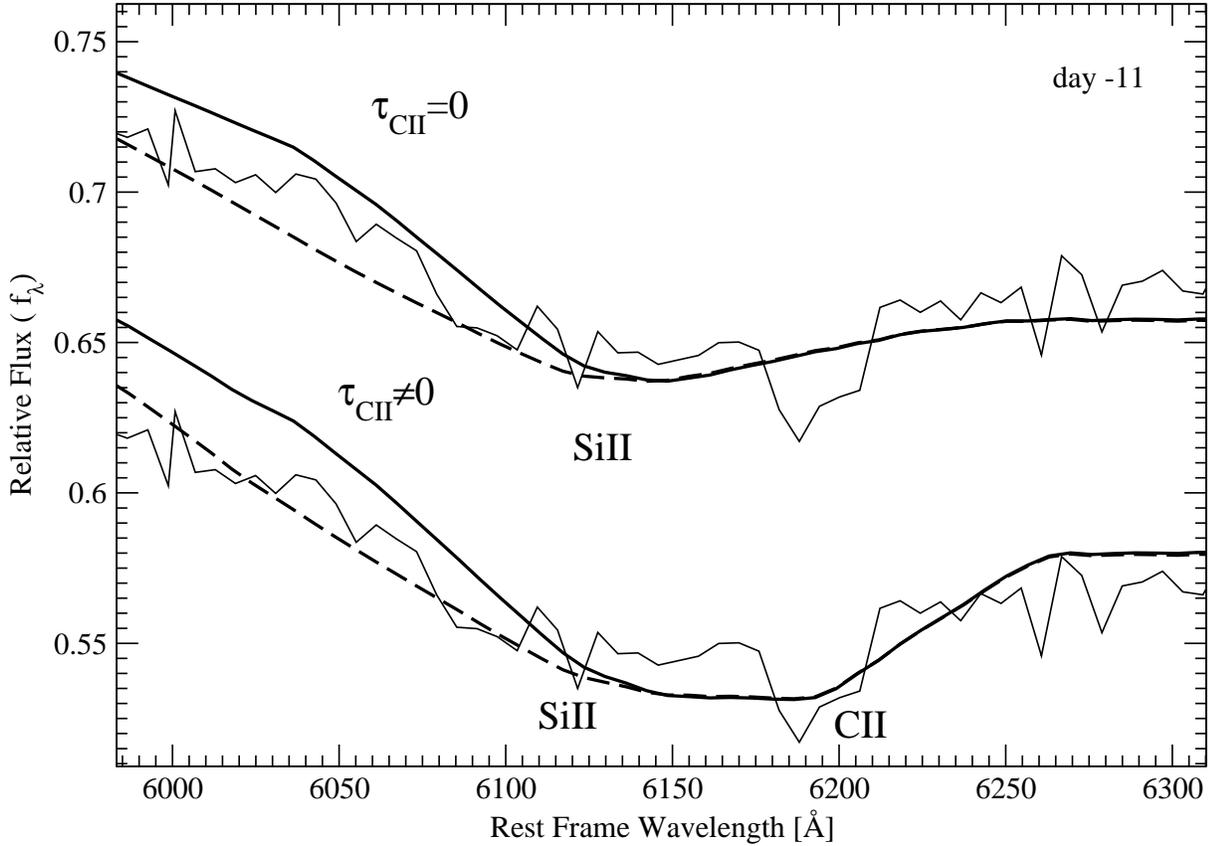}
  \caption{Synthetic spectra compared with SN~1999aa spectrum for day
  $-$11 around 6150~\AA. First model from the top: Solid lines,
  $\tau_{\rm CII}=0$ and data; Dashed lines, $\tau_{\rm CII}=0$ and
  Si~{\sc ii} with ${\rm v}_{max}=30 \times 10^{3}$km~s$^{-1}$. Second
  model from the top: Solid lines, $\tau_{\rm CII}\neq 0$ and data;
  Dashed lines, $\tau_{\rm CII}\neq 0$ and Si {\sc ii} with ${\rm
  v}_{max}=30 \times 10^{3}$km~s$^{-1}$. The continuum level has been
  shifted to match the data. Ions responsible for features in the
  synthetic spectrum are labeled. The position of the minimum changes
  by about 50~\AA\ when the the C~{\sc ii} component is included. The
  small notch around 6200~\AA\ is an artifact of the spectrum
  extraction and does not have to be considered a supernova feature.}
  \label{synow-11CII}
\end{figure*}

Guided by the low velocity value of Si~{\sc ii}~$6355$ and by the
similarity with SN~1991T we also introduced a HV C~{\sc ii}
(v$_{min}$=$19 \times 10^{3}$ km~s$^{-1}$) in order to try to improve
the description of the line profile at 6150~\AA.
Fig. \ref{synow-11CII} shows different SYNOW spectra compared with the
data in the 6150~\AA\ region. The small notch around 6200~\AA\ is an
artifact of the spectrum extraction and does not have to be considered
a supernova feature. The error spectrum on the wavelength region
6195-6210~\AA\ is 30$\%$ larger than that of neighboring regions thus,
the small notch does not have large statistical weight for any
quantitative measurements.  The continuum level of the models has been
shifted to match the data in this region.  The model at the top has no
C and includes only Si~{\sc ii}.  The absence of a C~{\sc ii}
component, i.e. considering only Si~{\sc ii}, places the minimum of
the feature to the blue side of the absorption at 6150~\AA\ and does
not reproduce completely the line profile. The parameter ${\rm v}_{\rm
min}$ is set equal to the photospheric velocity and thus has been
determined looking at all the lines together.  The dashed line
indicates the synthetic spectrum with the maximum velocity of Si~{\sc
ii} set to a value (${\rm v}_{\rm max}=30 \times 10^{3}$km~s$^{-1}$)
higher than in the later epochs, but which reproduces more accurately
the line profile.  The model at the bottom of Fig. \ref{synow-11CII}
includes both carbon and silicon.  Because of the introduction of
C~{\sc ii}, we would expect a small absorption (C~{\sc ii}
~$\lambda$7234) around 6800~\AA\ (close to a telluric line) which is
not convincingly visible.  It is not fully evident which of the two
models proposed matches better the data. The result of a $\chi^{2}$
analysis shows that they are equally plausible. However, the
introduction of C~{\sc ii} would solve the puzzle of the low velocity
value measured for Si~{\sc ii}~$\lambda$6355.  Note that later on (at
day $-$7, Fig. \ref{sn99aa.evo.ps}) the shape of the feature at
6150~\AA\ changes with the appearance of the first strong Si~{\sc ii}
~$\lambda$6355 line.

\begin{figure*}
\centering
 \includegraphics[width=16cm]{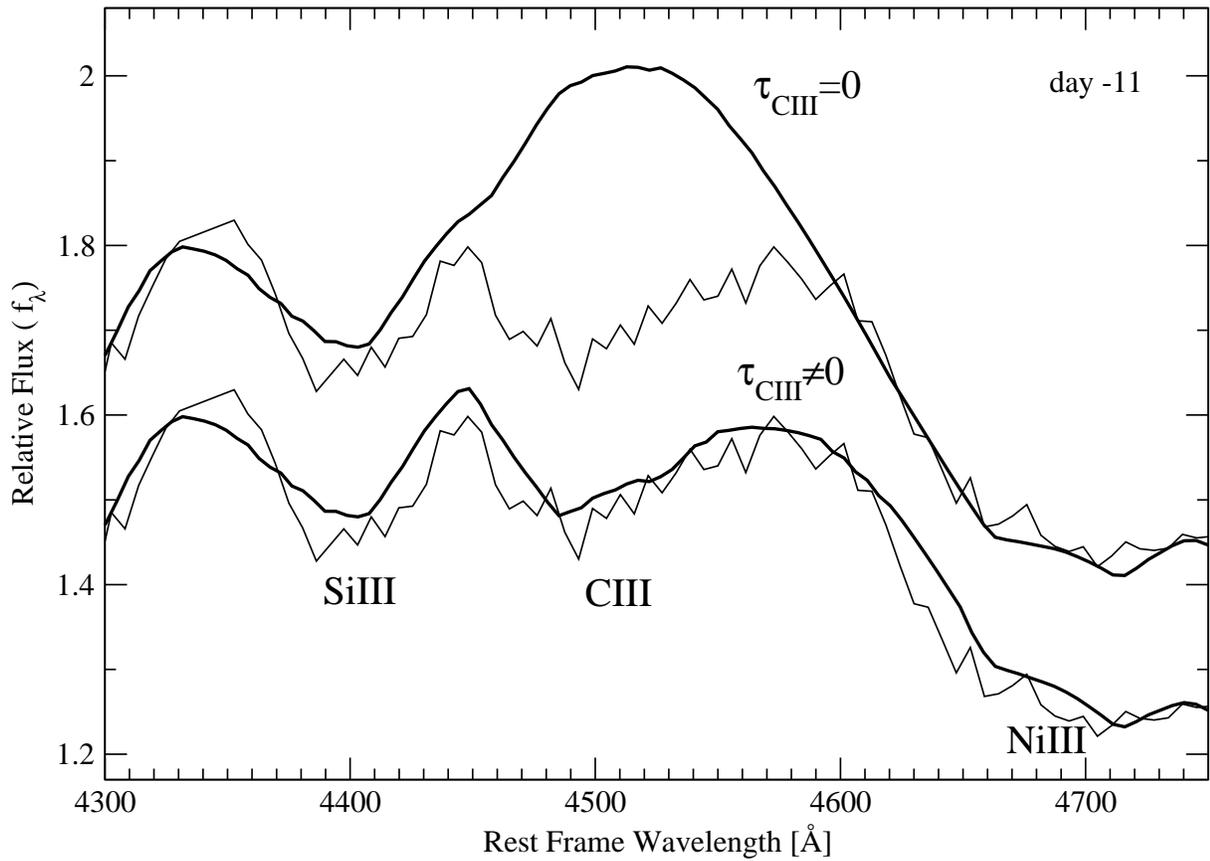}
  \caption{Synthetic spectra compared with SN~1999aa spectrum for day
  $-$11 around 4500~\AA. Solid lines from top to bottom: $\tau_{\rm C
  III}=0$ and data, $\tau_{\rm C III}\neq 0$ and data.}
  \label{synow-11CIII}
\end{figure*}

In our spectrum of SN~1999aa we find that the absorption on the red of
Si~{\sc iii}~$\lambda$4568 can be well matched by C~{\sc iii}
~$\lambda$4648.8 confined to low velocities. Its contribution is shown
in Fig.~\ref{synow-11CIII} where synthetic spectra with and without a
C~{\sc iii} component are compared with the observed spectrum. This
absorption is unusual for SNe~Ia and its identification is still
unclear. Three dimensional models indicate that unburned
doubly-ionized C can, in principle, be visible in the hot interior of
the supernova atmosphere \markcite{2003Sci...299...77G}({Gamezo} {et~al.} 2003). Later in the
spectral evolution, the identification of C could be difficult due to
increased blending of spectral lines. Further, we note that this
absorption feature is no longer visible in our spectrum at day $-$7
(see Fig. \ref{sn99aa.evo.ps}).

As an alternative to C~{\sc iii} and C~{\sc ii}, H could be used to
match the absorption at 4500~\AA\ and 6150~\AA\
\markcite{2002ApJ...580..374L}({Lentz} {et~al.} 2002).  We discuss these alternative
explanations in section \ref{carbonio}.

\subsection{Day 1 before maximum}
\begin{figure*}
\centering
   \includegraphics[width=16cm]{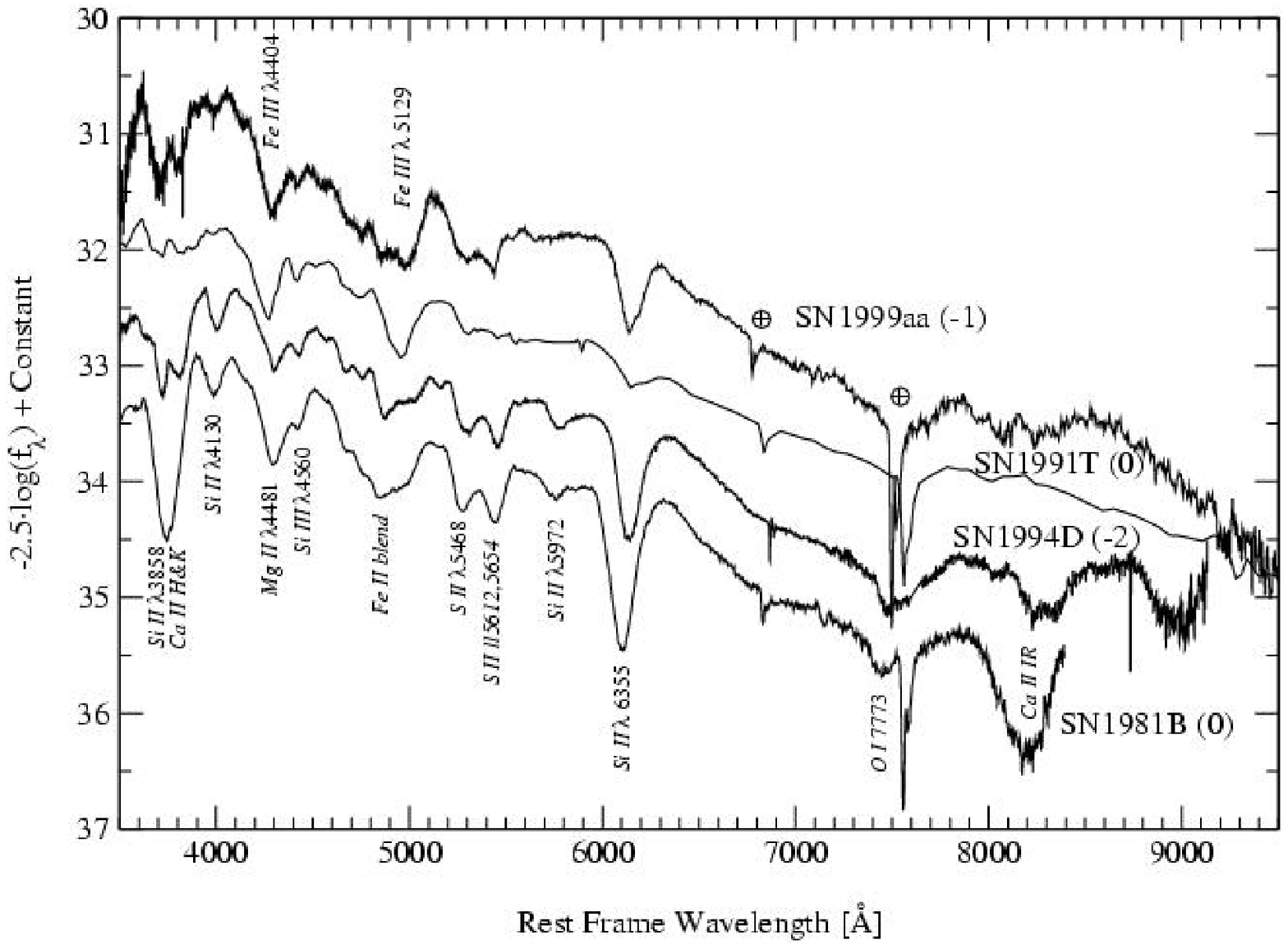}
 \caption{The $-$1 day spectrum of SN~1999aa together with those of
  SN~1991T, SN~1981B and SN~1994D from
  \markcite{1992ApJ...397..304J,1983ApJ...270..123B,1996MNRAS.278..111P}{Jeffery}
  {et~al.} (1992); {Branch} {et~al.} (1983); {Patat} {et~al.}
  (1996). Each spectrum is labeled with the phase (days since B
  maximum). The $\oplus$ symbol marks atmospheric absorptions. }
  \label{comp0.ps}
\end{figure*}

The spectrum of SN~1999aa around maximum light is shown in
Fig.~\ref{comp0.ps} together with those of SN~1991T, SN~1981B and
SN~1994D.  SN~1999aa resembles that of normal SNe~Ia, while SN~1991T
is still dominated by Fe~{\sc iii} lines with very weak Si~{\sc ii}
and Ca~H\&K lines at this epoch. The Ca~{\sc ii}~H\&K region of
SN~1999aa shows a characteristic split where the red component is
gaining strength. S~{\sc ii} appears at this epoch showing the typical
`W' shaped feature at $\lambda\lambda$5454,5606~\AA\ that
SN~1991T-like SNe~Ia do not usually show.  Si~{\sc ii}~$\lambda$6355
is visible but fainter and redder than in normal SNe~Ia. If the
evolution from C~{\sc ii} to Si~{\sc ii} proposed by
\markcite{1999MNRAS.304...67F}{Fisher} {et~al.} (1999) is confirmed, SN~1991T at this epoch shows
a weaker Si~{\sc ii}~$\lambda$6355 absorption than SN~1999aa.  This
suggests that the amount of Si~{\sc ii} present in SN~1999aa and/or
the atmospheric temperature could be between that of normals and
SN~1991T-like SNe~Ia. The Ca~{\sc ii} IR triplet around 8000~\AA\ has
two weak features not present in SN~1981B but visible in SN~1994D. All
the other major features are the same as those seen in normal SNe~Ia.

\subsection*{Synthetic spectrum}
\begin{figure*}
\centering
   \includegraphics[width=16cm]{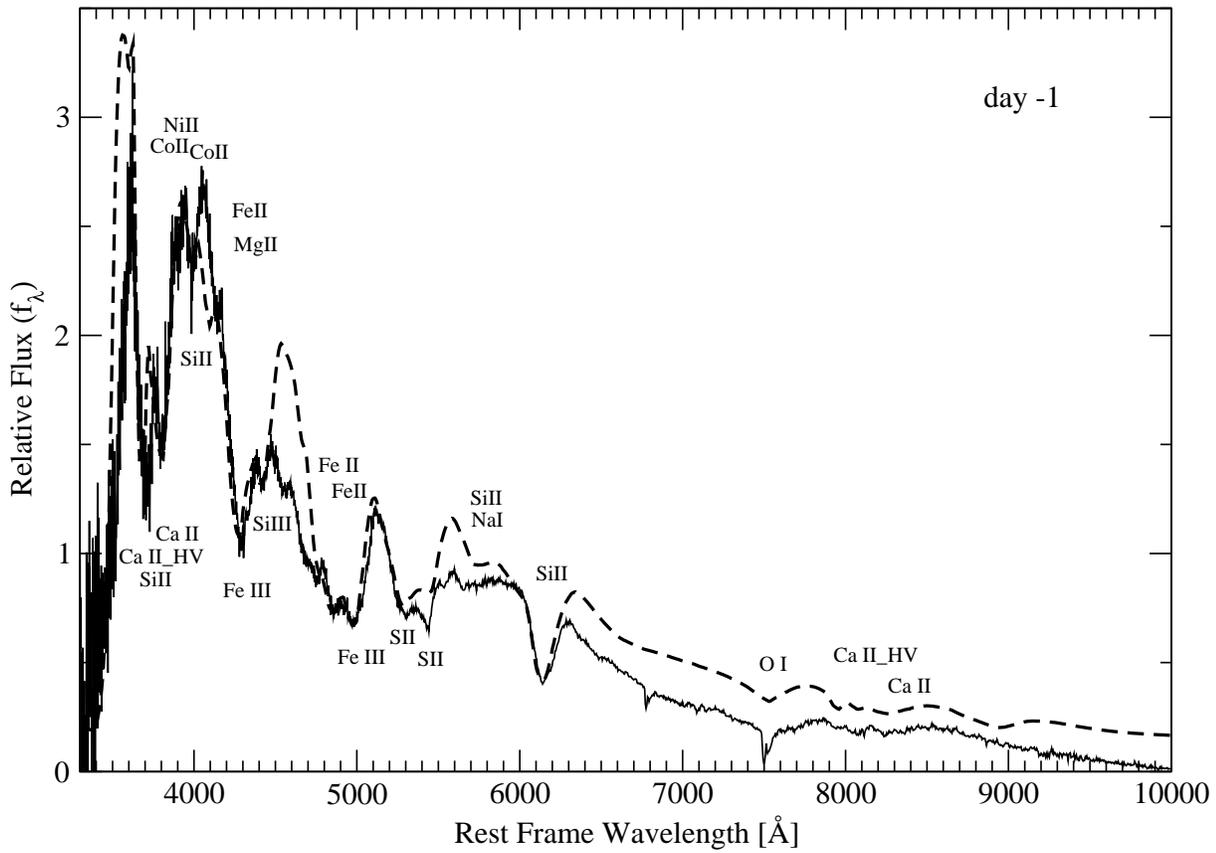}
  \caption{Synthetic spectrum compared with SN~1999aa spectrum for day
  $-$1. SYNOW parameters used are presented in Table
  \ref{table-1}. The region around 3800~\AA\ and 8000~\AA\ are
  enlarged in Figs. \ref{synow-1ca} to highlight the contribution of
  Ca~{\sc ii}~HV.  Above 6200~\AA\ the black body assumption fails in
  reproducing the correct flux values but it does not affect the
  identification of the lines. Ions responsible for features in the
  synthetic spectrum are labeled.}
  \label{synow-1}
\end{figure*}

The synthetic spectrum for day $-$1 is shown Fig. \ref{synow-1}. As in
the previous epoch, Si~{\sc ii} and Si~{\sc iii} have been introduced
with a low maximum velocity in order to match the observed wavelengths
of the minima; see Table \ref{table-1} for details. The S~{\sc ii}, Mg
{\sc ii}, Co~{\sc ii} and Fe~{\sc ii} lines are well reproduced in
detached layers.

\begin{figure*}
\centering
  \includegraphics[width=16cm] {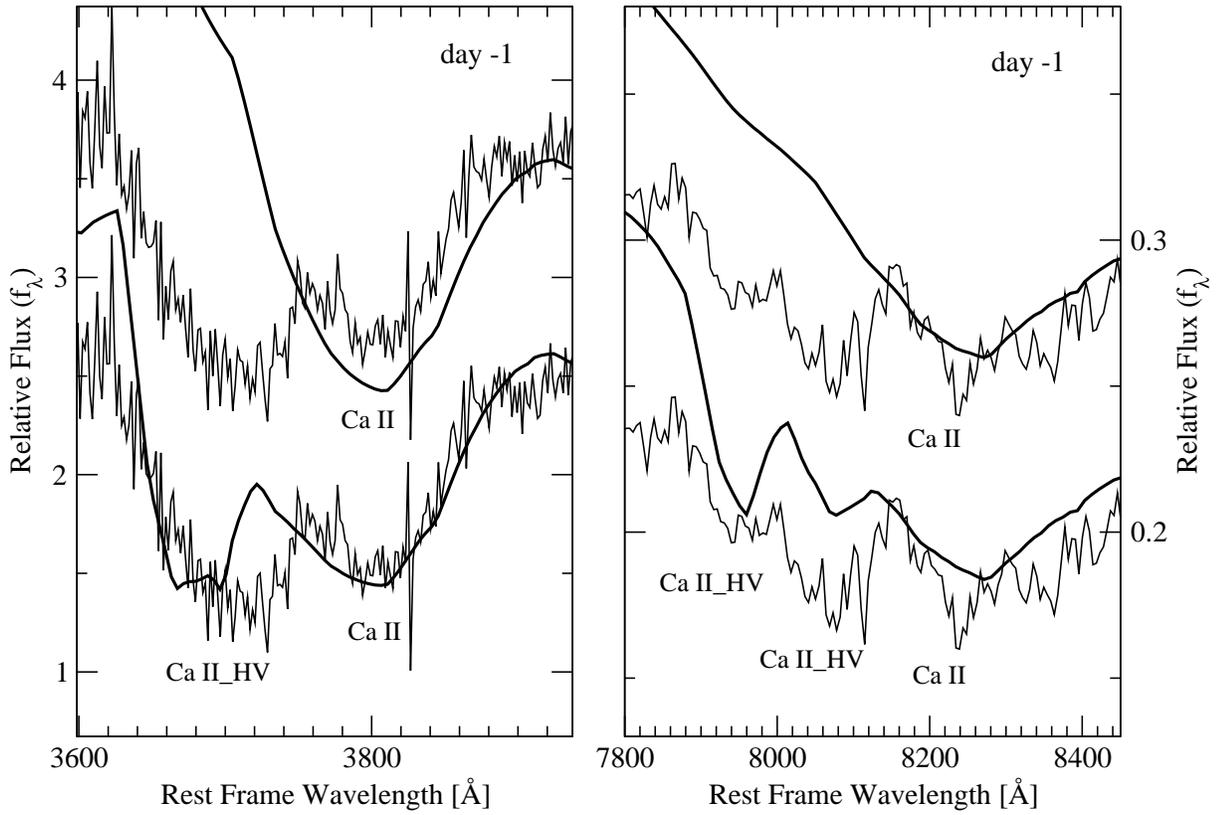}
  \caption{Synthetic spectrum compared with SN~1999aa spectrum for day
  $-$1 around 3800~\AA\ (left panel) and 8000~\AA\ (right
  panel). First model from the top: Solid lines, $\tau_{\rm Ca
  II~HV}=0$ and data; Second model from the top: Solid lines,
  $\tau_{\rm Ca II~HV}\neq0$ and data; The continuum level on the
  right panel has been shifted to match the data. Ions responsible for
  features in the synthetic spectrum are labeled.}
  \label{synow-1ca}
\end{figure*}

The Ca H\&K feature is formed by a blend of the two dominant
components of Ca~{\sc ii} and weaker contributions from Si~{\sc ii},
Co {\sc ii} and Ni~{\sc ii}.  The double component of Ca~{\sc ii} is
required at this epoch both in the H\&K and IR triplet regions
(respectively left and right panel in Fig \ref{synow-1ca}). The split
seen at the location of Ca H\&K has been reproduced with Si~{\sc
ii}~$\lambda$3858 or Si~{\sc iii}~$\lambda$3801 in NLTE simulations
\markcite{1997ApJ...485..812N,2000ApJ...530..966L}({Nugent} {et~al.} 1997; {Lentz} {et~al.} 2000) but with SYNOW, the
only way to account for the blue minimum of Ca H\&K is to add a high
velocity component (Ca~{\sc ii}~HV) \markcite{1999ApJ...525..881H}({Hatano} {et~al.} 1999a).
Ca~{\sc ii}~HV with ${\rm v}_{\rm min}=21000$~km~s$^{-1}$ also matches
the two small absorptions in the blue part of the Ca~{\sc ii} IR
triplet. In the right panel of Fig.~\ref{synow-1ca} the black body
continuum has been shifted to the level of the observed spectrum
around 8000~\AA\ to show the match in velocity. The correct relative
intensity of the two HV Ca~{\sc ii} IR triplet features and
the correct line profile are better reproduced by modeling the 3D
geometry of the HV clump
\markcite{2003ApJ...593..788K,Thomas:2003ip}({Kasen} {et~al.} 2003; Thomas {et~al.} 2003). In this work we are only
interested in identifying the velocity range at which this component
is present.

The absorption features we tentatively identified as C~{\sc ii} and
C~{\sc iii} in the early spectrum are no longer detectable. The broader
and stronger Si~{\sc ii}~$\lambda$6355 and Si~{\sc iii}~$\lambda$5051
now dominate the wavelength ranges at which these lines were present.

As seen in Fig. \ref{synow-1}, the synthetic spectrum fails 
to describe the data in several wavelength regions: around
 4600~\AA, 5500~\AA\ and beyond 6200~\AA. In similar analyses with
SYNOW, attempts were made to improve the match around  4600~\AA\  
by adding a high-velocity component of Fe~{\sc ii}. For SN~1999aa,
this approach introduced mismatches in other wavelength regions.
The disagreement, as well as that around 5500~\AA, 
 could also be (partly) explained by the inaccuracy of
the blackbody continuum, which is clearly showing a mismatch
for the flux redward of 6200~\AA.

As for the spectrum at day $-$11, the black body continuum fails to
produce the correct flux redward of 6200~\AA\ but this does not affect
the reliability of the line profile reproduction.

Despite attempting several different parameter combinations, the
synthetic spectrum shows significant disagreement with the data at
4600~\AA\ on the red side of S~{\sc ii}.

\subsection{Day 5 after maximum}
\begin{figure*}
\centering
   \includegraphics[width=16cm]{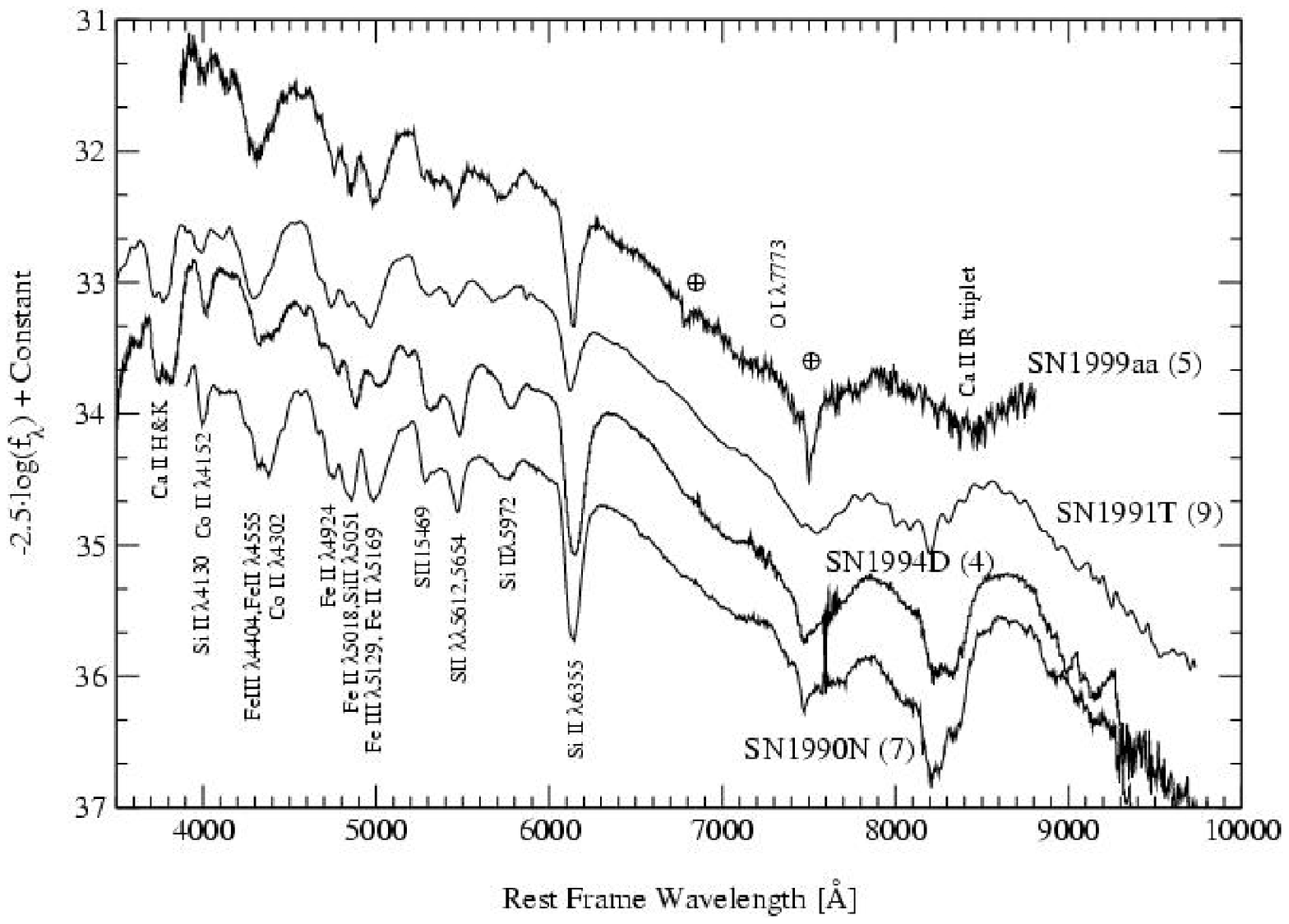}
  \caption{The +5 day spectrum of SN~1999aa together with those of
  SN~1991T, SN~1990N and SN~1994D from
  \markcite{1992ApJ...384L..15F,1991ApJ...371L..23L,1996MNRAS.278..111P}{Filippenko}
  {et~al.} (1992); {Leibundgut} {et~al.} (1991); {Patat} {et~al.}
  (1996). Each spectrum is labeled with the phase (days since B
  maximum). The $\oplus$ symbol marks atmospheric absorptions.}
  \label{comp6.ps}
\end{figure*}

The spectrum of SN~1999aa around one week after maximum, is shown in
Fig.~\ref{comp6.ps} and compared with those of SN~1991T, SN~1994D and
SN~1990N.  The spectrum of SN~1999aa is quite normal, now also showing
the blue component of the Si~{\sc ii} duo in the 5700-6200~\AA\
region.  SN~1991T still exhibits weaker Si~{\sc ii} and S~{\sc ii}
lines at this epoch.  In normal SNe, absorption features due to
elements such as Co~{\sc ii}, Fe~{\sc ii}, Ca~{\sc ii}~IR triplet and
O~{\sc i}~$\lambda$7773 get stronger at this time.  From
Fig.~\ref{comp6.ps} and from the spectra of other SN~1991T-like
supernovae shown in \markcite{1999AJ....117.2709L,2001PASP..113.1178L}{Li} {et~al.} (1999, 2001a),
it is clear that the general trend for these objects at this epoch is
to still have very weak intermediate mass element components and a
persistent dominant presence of doubly ionized Fe.  Furthermore,
SN~1997br and SN~2000cx
\markcite{1999AJ....117.2709L,2001PASP..113.1178L}({Li} {et~al.} 1999, 2001a) already have a large
contribution from low ionization iron-peak elements (more prominent
than in Branch-normal SNe), with SN~1997br more advanced in this
transition.  This suggests a very interesting sequence among these
objects in which the duration in the phase of the intermediate mass
elements is shortest for SN~1997br, and longest (and thus more similar
to normal SNe) for SN~1999aa, with SN~1991T placed somewhere in the
middle. For Branch-normal supernovae, this phase extends from well
before maximum to a couple of weeks after. In the case of these
peculiar objects it seems instead to vary in duration and phase during
the time evolution. This is probably due to the ratio of Fe-peak to
IME abundances, their radial distribution and temperature
profile. Further studies are needed to quantify these differences.

\subsection*{Synthetic spectrum}
In the synthetic spectrum for day 5 after maximum, shown in
Fig. \ref{synow+5}, doubly ionized Si and Fe have lower optical depth
than in the previous epoch analyzed; see Table \ref{table+5}. Fe~{\sc iii}
and Si~{\sc iii} are now blended, forming a deep trough at $4400$~\AA\
together with Co~{\sc ii} and Mg~{\sc ii}.
We also note a mismatch between the data and the synthetic spectrum
around $4600$~\AA.

In order to reproduce the three distinct minima below $5000$~\AA, the
Fe~{\sc ii} layer has to be detached, with a similar velocity range as
in day $-$1, otherwise the lines tend to blend together forming a
single broad feature.

At this epoch, Si~{\sc ii} and Si~{\sc iii} also become detached,
as did S {\sc ii} and Mg~{\sc ii} in the previous epoch. Now they all
appear confined above $10\times 10^{3}$km~s$^{-1}$. This is a probable sign
that complete silicon burning has stopped at this velocity.

The feature at $5700$~\AA\ is a blend of Na~{\sc i} with a weak
Si~{\sc ii} component. The Si~{\sc ii} component is weak because of
the low excitation temperature used; a higher temperature would create
a redder feature than the one seen in our data.

Unfortunately at this epoch our spectrum does not cover the region
below $3900$~\AA\ and the Ca~IR triplet is too weak to fully determine
whether there is still need for a HV component.

\subsection{Day 14 after maximum}
\begin{figure*}
\centering
   \includegraphics[width=16cm]{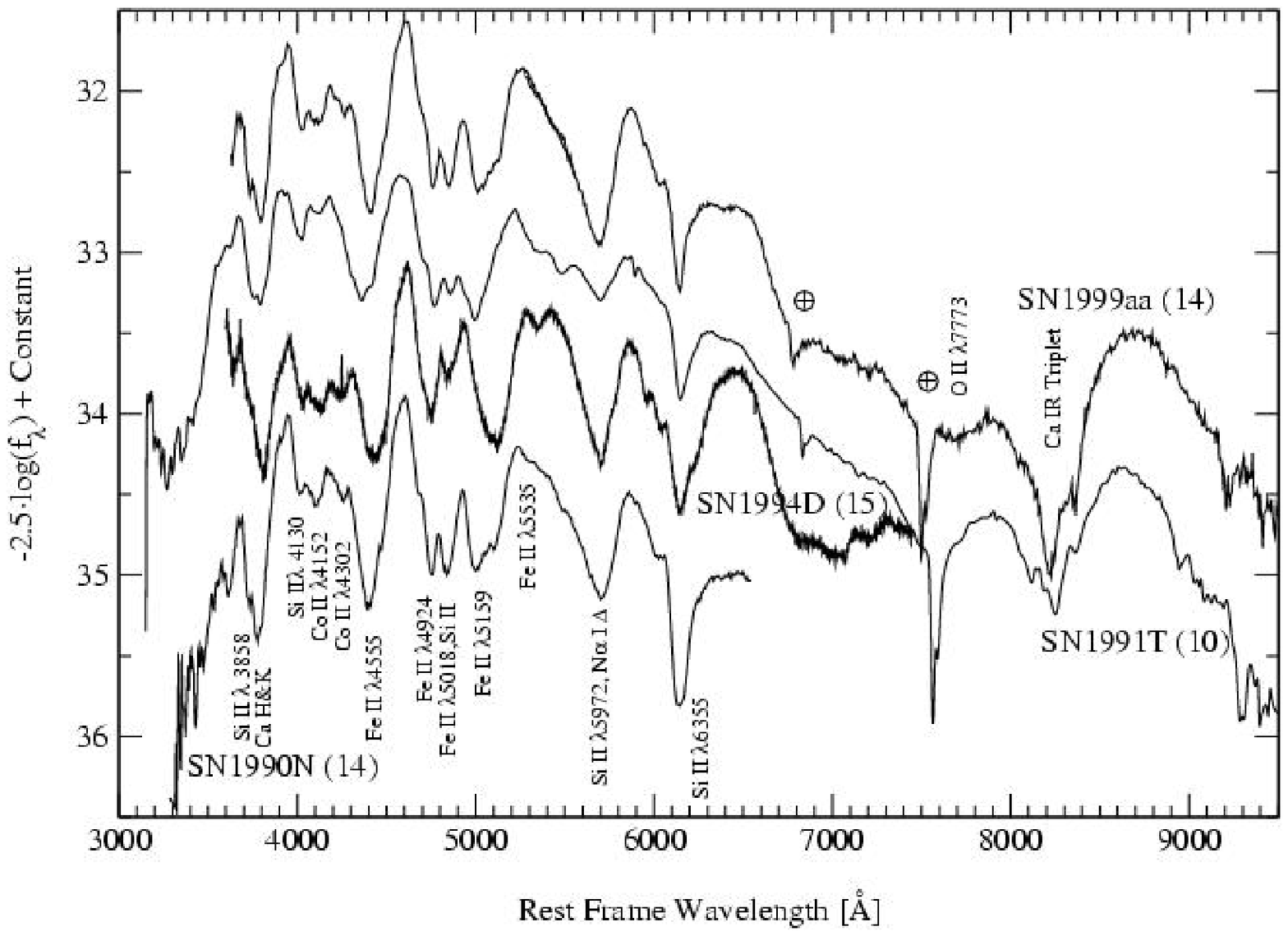}
\caption{The +14 day spectrum of SN~1999aa together with those of
  SN~1991T, SN~1990N and SN~1994D from
  \markcite{1992AJ....103.1632P,1991ApJ...371L..23L,1993A&A...269..423M}{Phillips}
  {et~al.} (1992); {Leibundgut} {et~al.} (1991); {Mazzali} {et~al.}
  (1993). Each spectrum is labeled with the phase (days since B
  maximum). The $\oplus$ symbol marks atmospheric absorptions.}
  \label{comp14.ps}
\end{figure*}

Our spectrum of SN~1999aa at day 14 is shown in Fig.~\ref{comp14.ps}
together with those of SN~1991T, SN~1994D and SN~1990N.  SN~1999aa is
now completely indistinguishable from a normal supernova, such as
SN~1990N, with the possible exception of a slightly weaker Si~{\sc
ii}~$\lambda$6355. This is to be expected, since small differences are
visible even among so called normals. In comparison SN~1991T is still
completing its intermediate mass elements phase; S~{\sc ii} has not
yet blended with Na~{\sc i} as in normal objects. The overall
contribution of S~{\sc ii}, Si~{\sc ii} and Ca~{\sc ii} is
consistently weaker than in typical type Ia.

\subsection*{Synthetic spectrum}
In the synthetic spectrum at day +14, shown in Fig. \ref{synow+14},
most of the lines can be identified as Fe~{\sc ii} (Table \ref{table+14}).
As in the previous synthetic spectrum, in order to reproduce the three
different minima in the Fe~{\sc ii} blend region (4800~\AA) we use a
minimum velocity higher than that of the photosphere (Table
\ref{table+14}).

The deep line at 5700~\AA\ is dominated by Na~{\sc i}, and its shape
can be reproduced if Na~{\sc i} is introduced with higher ${\rm
v_{e}}$ than that used for the other ions.

During the entire spectral evolution, the flux level and shape in the
region between $6500$~\AA\ and $8000$~\AA\ has become increasingly
more difficult to match. At this epoch, the discrepancy between data
and synthetic spectrum is fairly evident. This suggests an increasing
inaccuracy of the assumption of an underlying black body continuum.

 \subsection{Early Nebular Phase Spectra}
\begin{figure*}
\centering
   \includegraphics[width=16cm]{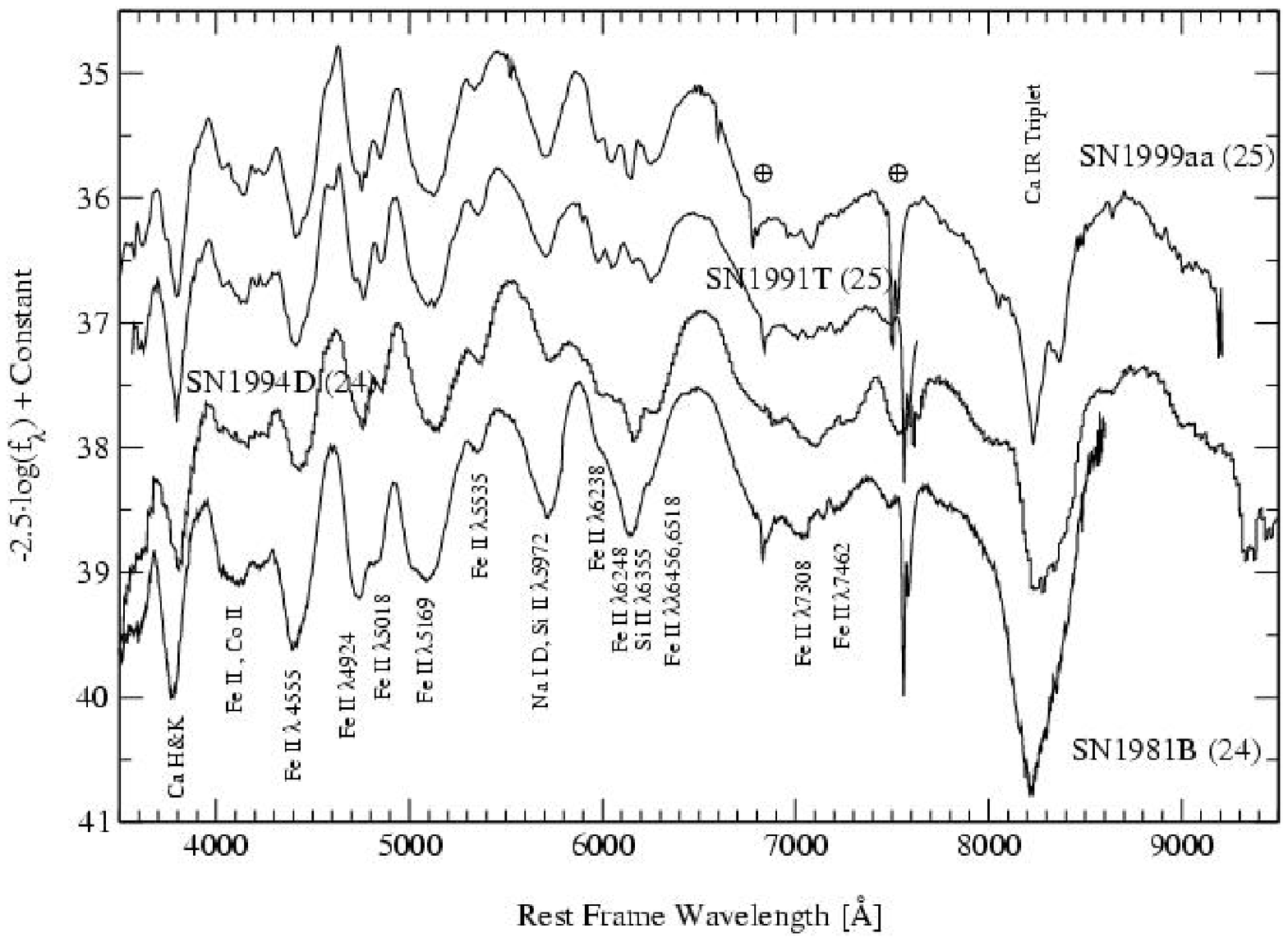}
  \caption{The +25 day spectrum of SN~1999aa together with those of
  SN~1991T, SN~1981B and SN~1994D from
  \markcite{1992AJ....103.1632P,1983ApJ...270..123B,1996MNRAS.278..111P}{Phillips}
  {et~al.} (1992); {Branch} {et~al.} (1983); {Patat} {et~al.}
  (1996). Each spectrum is labeled with the phase (days since B
  maximum). The $\oplus$ symbol marks atmospheric absorptions.}
  \label{comp25.ps}
\end{figure*}

The day +25 days of SN~1999aa is shown in Fig. \ref{comp25.ps}.  For
comparison SN~1991T, SN~1994D and SN~1981B are also shown.  The main
differences in the observed spectra lie in the region around 6000~\AA,
and are probably due to different Si abundances among the different
SNe Ia. SN~1991T shows the Fe~{\sc
ii}~$\lambda\lambda$6238,6246,6456,6516 lines and just a very weak
Si~{\sc ii} line in the central part of that wavelength region. In
SN~1999aa, Si~{\sc ii} is more evident, and it becomes even stronger
in SN~1994D and SN~1981B. The trough near 5700~\AA\ due to Si~{\sc
ii}~$\lambda$5972 and Na~{\sc i} D is strong in all the SNe. At the
red end of the spectra, the four SNe show the typical Ca~{\sc ii}~IR
triplet as a very deep and broad absorption. SN~1999aa shows a second
minimum in the red part of this feature not firmly identified, but
possibly consistent with an absorption feature due to Co~{\sc ii}.
 
\begin{figure*}
\centering
   \includegraphics[width=16cm]{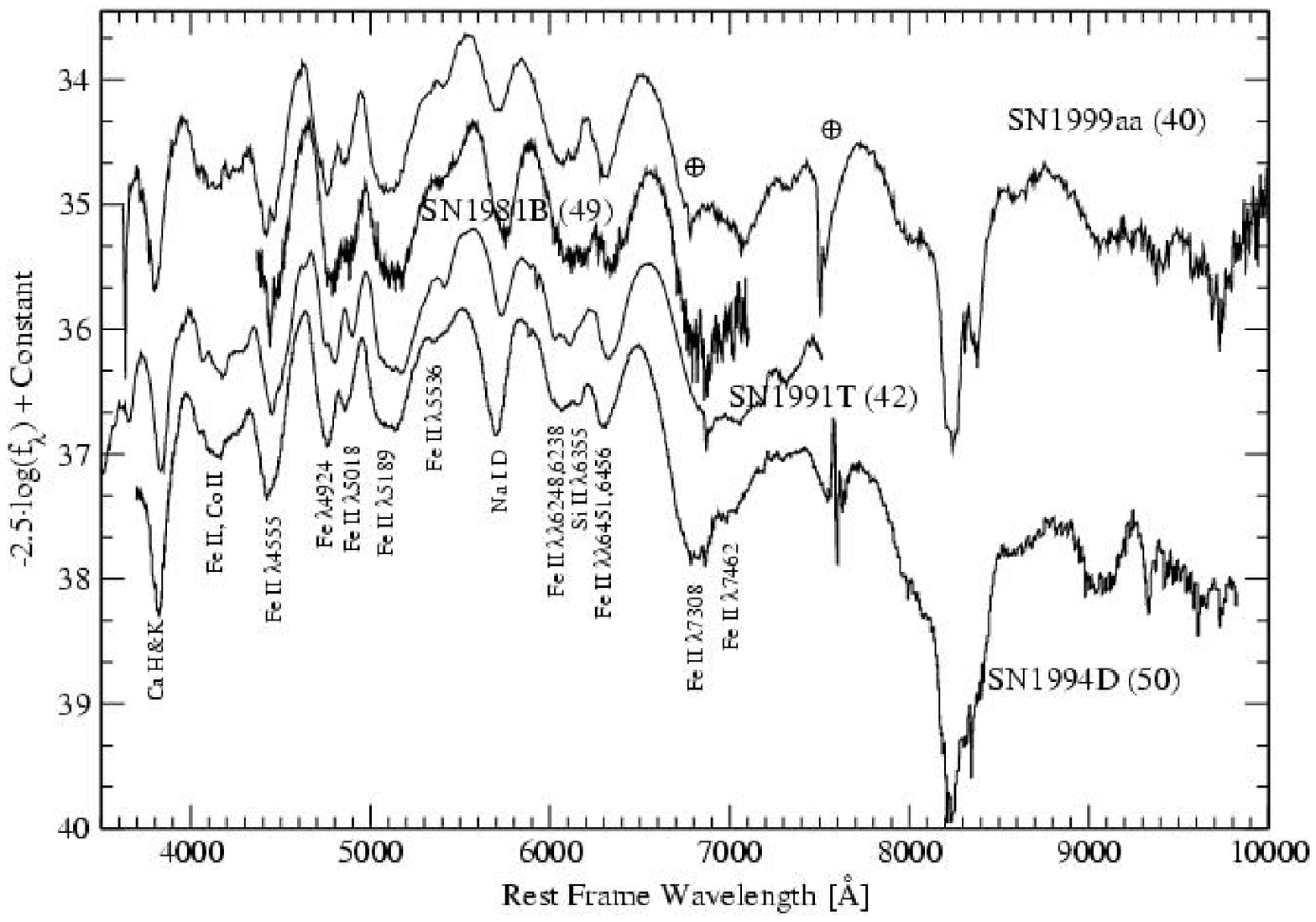}
  \caption{The +40 day spectrum of SN~1999aa together with those of
  SN~1991T, SN~1981B and SN~1994D from
  \markcite{1996AJ....112.2094G,1983ApJ...270..123B,1996MNRAS.278..111P}{Gomez},
  {Lopez}, \& {Sanchez} (1996); {Branch} {et~al.} (1983); {Patat}
  {et~al.} (1996). Each spectrum is labeled with the phase (days since
  B maximum). The $\oplus$ symbol marks atmospheric absorptions.}
  \label{comp47.eps}
\end{figure*}

The spectrum of SN~1999aa at day +40 is shown in Fig.~\ref{comp47.eps}
and compared with those of SN~1991T, SN~1994D and SN~1981B.  At this
stage the spectra are all dominated by Fe~{\sc ii} and Co~{\sc ii}
lines formed mainly in the deep layers of the atmosphere.  Only small
differences are now visible in the depths of some lines.

\subsection*{Synthetic spectrum}
At day +40 the supernova has started entering the so called
nebular phase and SYNOW's assumption of a sharp photosphere becomes
less physically realistic.  Even though SYNOW was designed only to
reproduce spectra in the photospheric phase, several authors have
drawn conclusions based on its matches to nebular spectra (see
e.g. \markcite{1999MNRAS.304...67F}{Fisher} {et~al.} (1999)).  In the interest of comparison with
their work, we also present the synthetic spectrum for this epoch.

\begin{figure*}
\centering 
 \includegraphics[width=16cm]{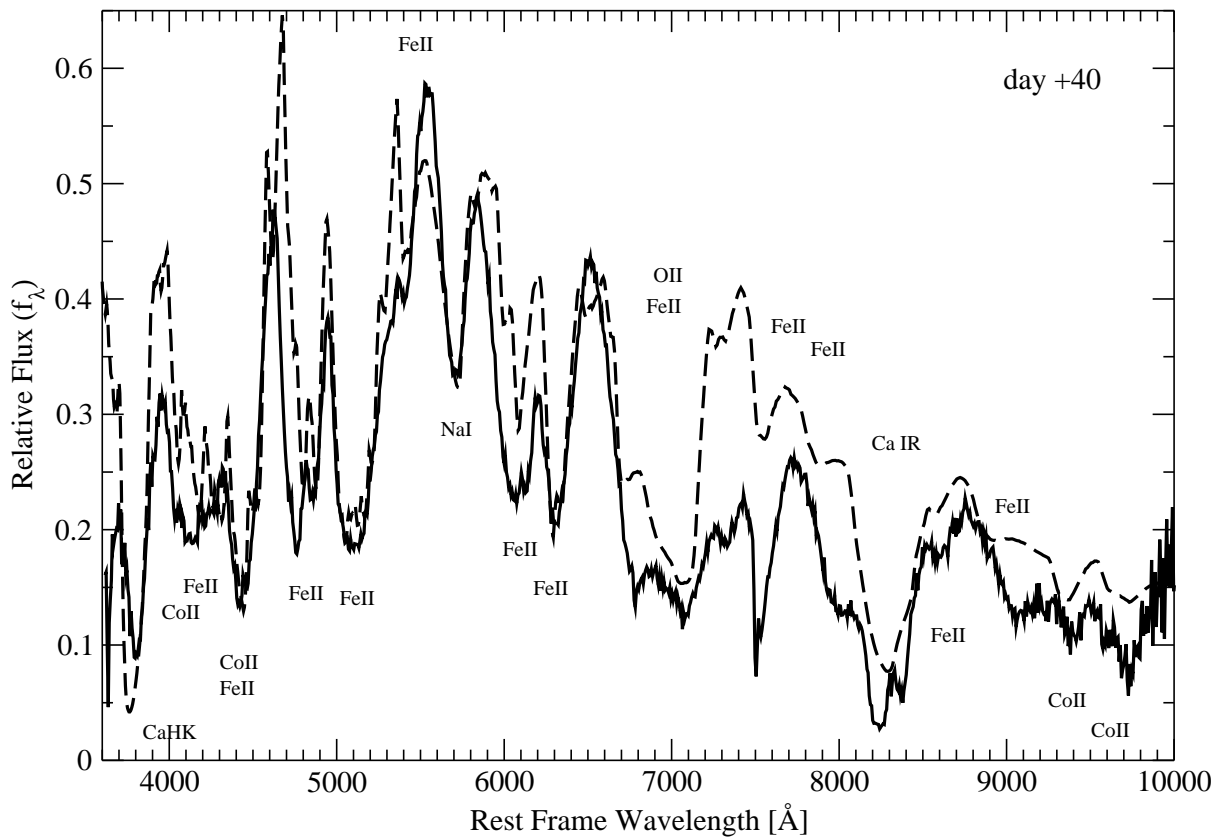}
  \caption{Synthetic spectrum compared with SN~1999aa spectrum for day
  +40. SYNOW parameters used are presented in Table \ref{table+40}. A
  discontinuity at 10000 km~s$^{-1}$ has been introduced in the Co and
  Fe optical depths, indicating the possible iron-peak core
  limit. Ions responsible for features in the synthetic spectrum are
  labeled. }
  \label{synow+40}
\end{figure*}

With the photosphere now into the iron-peak core, at this epoch the
supernova spectrum is mainly formed by Fe~{\sc ii} and Co~{\sc ii};
see Fig \ref{synow+40}.  The only signs of IMEs are Ca~{\sc ii} H\&K
and the Ca~IR triplet. As proposed by \markcite{1999MNRAS.304...67F}{Fisher} {et~al.} (1999) for
SN~1991T, the strong and wide feature near $7000$~\AA\ can be
reproduced by forbidden lines of [O~{\sc ii}]
($\lambda\lambda$7320,7330) with a high ${\rm v}_{\rm e}$. Note that
in SYNOW this line is also treated with a resonance scattering source
function. For the details of the composition, see Table
\ref{table+40}.  Special care was taken in the way the components of
Fe~{\sc ii} and Co~{\sc ii} have been introduced. At this epoch, the
spectrum is formed in the Fe-peak core but still a part of the outer
atmosphere makes a contribution. The optical depth of these elements
in the two regions can be different.

\markcite{1998ApJ...499L..49M}{Mazzali} {et~al.} (1998) demonstrated
that the expansion velocity of the Fe-peak core and the luminosity of
SNe~Ia are correlated. Slow decliner supernovae, such as SN~1991T, are
expected to have a higher Fe-peak core limit velocity and thus a
larger region dominated by nuclear statistical equilibrium.  In this
respect, it is interesting to attempt the identification of the
iron-peak core limit by means of SYNOW synthetic spectra.
\markcite{1999MNRAS.304...67F}{Fisher} {et~al.} (1999) and
\markcite{2002NewA....7..441H}{Hatano} {et~al.} (2002) introduced an
optical depth discontinuity for Co~{\sc ii} and Fe~{\sc ii} at 10000
km~s$^{-1}$ and 12500 km~s$^{-1}$ for their synthetic spectra of
SN~1991T and SN~1997br respectively. The velocity at which this change
occurs can be thought of as the iron peak core limit. We tried to
reproduce this for SN~1999aa. The best match was achieved with a
discontinuity at $10000$~km~s$^{-1}$.

\section{ Expansion Velocities}
\label{sec_velocities}
\begin{figure*}
\centering 
 \includegraphics[width=16cm]{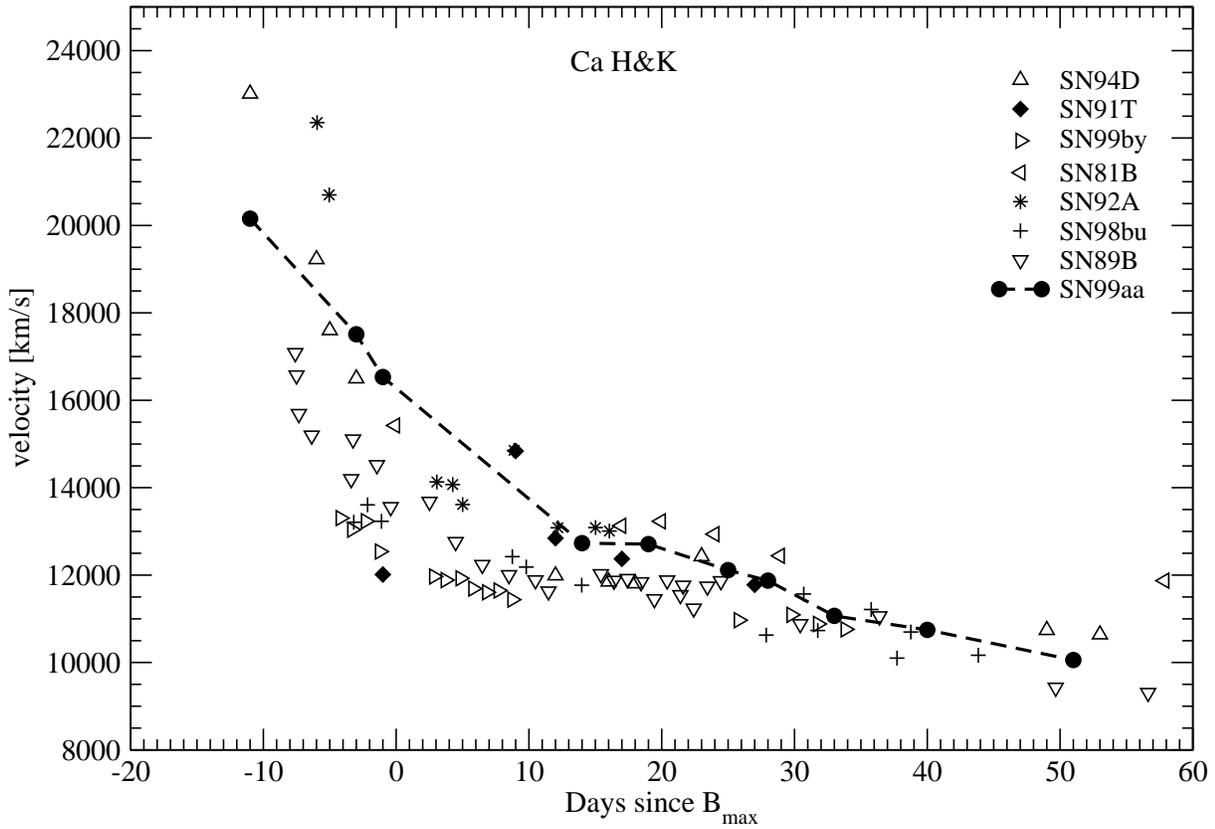}
  \caption{Doppler shift of the Ca H\&K. The values for SN~1999aa are
  compared with those for other SNe~Ia taken from
  \markcite{1996AAS...189.4510W,Garnavich:2001vx,1993ApJ...415..589K,1996MNRAS.278..111P,1999ApJS..125...73J}{Wang} {et~al.} (1996); Garnavich {et~al.} (2001); {Kirshner} {et~al.} (1993); {Patat} {et~al.} (1996); {Jha} {et~al.} (1999) and references
  therein.}
  \label{CaII.eps}
\end{figure*}

\begin{figure*}
  \centering
 \includegraphics[width=16cm]{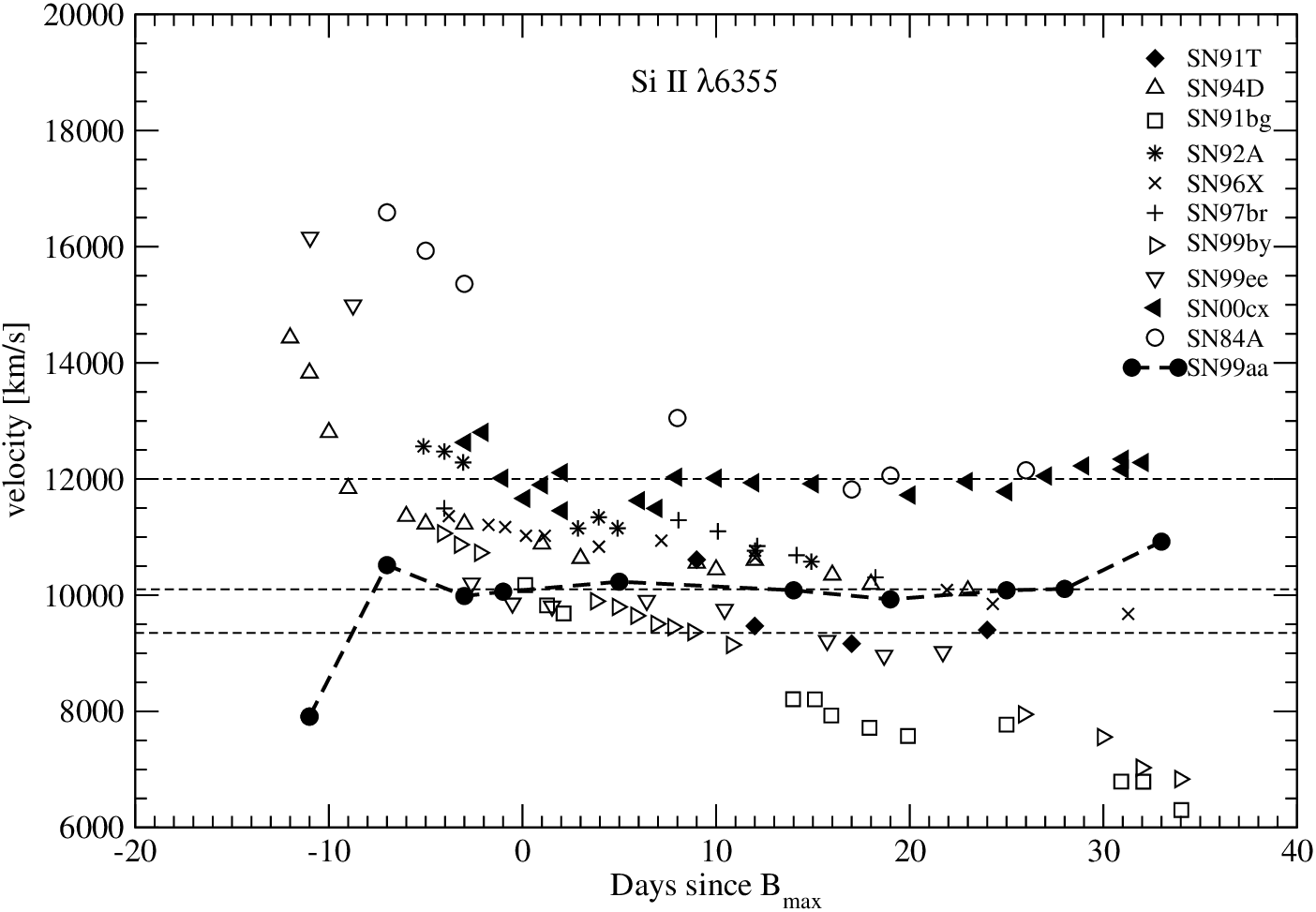}
  \caption{ Doppler shift of the Si~{\sc ii} $\lambda$6355. The values for
  SN~1999aa are compared with those for other SNe~Ia taken from
  \markcite{1999AJ....117.2709L,2001PASP..113.1178L,Garnavich:2001vx,2001MNRAS.321..254S}{Li} {et~al.} (1999, 2001a); Garnavich {et~al.} (2001); {Salvo} {et~al.} (2001) and references therein. The
  low value of SN~1999aa at day $-$11 is probably due to C~{\sc ii} or H
  contamination. SN~2000cx, SN~1999aa and SN~1991T maintain a constant
  velocity of 12000~km~s$^{-1}$, 10100~km~s$^{-1}$ and 9400~km~s$^{-1}$ (dashed lines)
  respectively from maximum light to after day 20.}
  \label{siII635_vel.eps}
\end{figure*}

\begin{figure*}
  \centering
 \includegraphics[width=16cm]{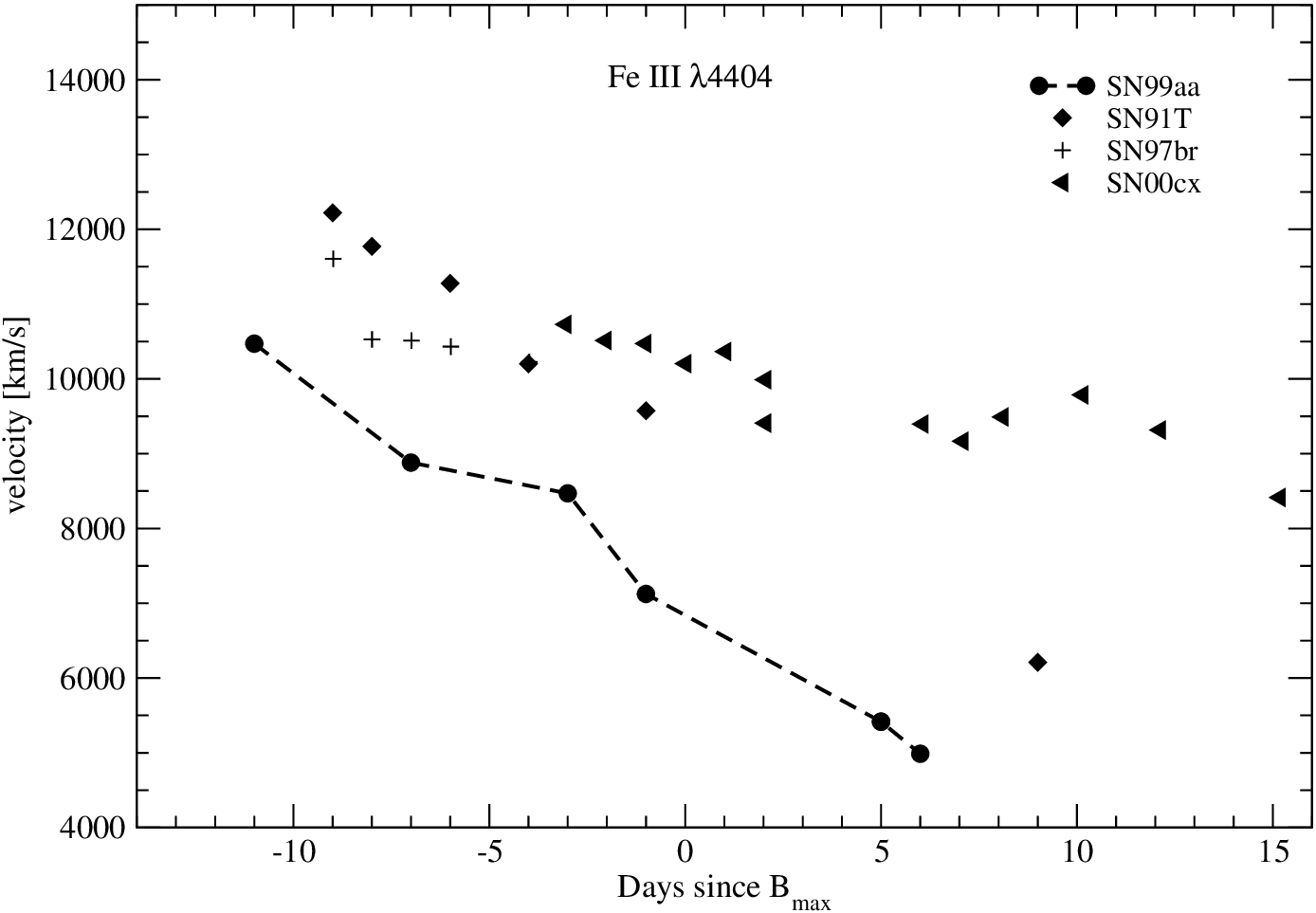}
  \caption{Doppler shift of the Fe~{\sc iii} $\lambda$4404 . The
  values for SN~1999aa are compared with those for other SNe~Ia taken
  from \markcite{1999AJ....117.2709L,2001PASP..113.1178L}{Li} {et~al.}
  (1999, 2001a) and references therein.  }
  \label{FeIII_vel.eps}
\end{figure*}

\begin{figure*}
  \centering
 \includegraphics[width=16cm]{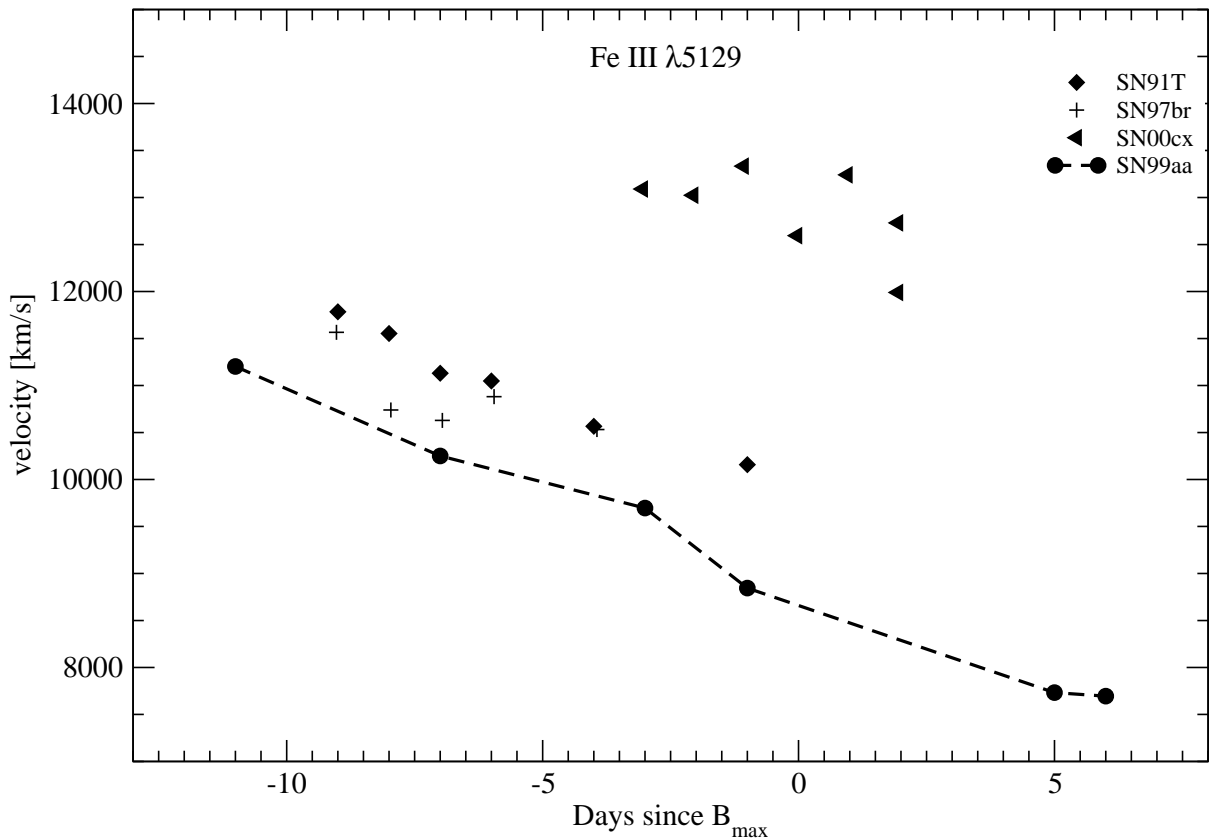}
  \caption{Doppler shift of the Fe~{\sc iii} $\lambda$5129 . The values for
  SN~1999aa are compared with those for other SNe~Ia  taken from
  \markcite{1999AJ....117.2709L,2001PASP..113.1178L}{Li} {et~al.} (1999, 2001a) and references therein.}
  \label{FeIII1_vel.eps}
\end{figure*}

The expansion velocities as computed from fits to the minima of the
spectral lines can provide help in investigating the physics of the
supernova explosion. Weighted fits of the observed minima were
performed using a non-linear Marquardt-Levenberg minimization
procedure \markcite{fitprocedure}({Marquardt} 1963) applied to a Gaussian profile
model. The minimum of the line was considered to be the center of the
Gaussian and the fit uncertainty its statistical error. This
statistical error is usually of the order of a few km~s$^{-1}$, so is
not shown on the graphs. In general, the fit was performed on the
entire absorption of the P-Cygni profile when this was well reproduced
by a Gaussian model, e.g. when there was a low level of contamination
by other lines. In the contaminated cases the fit was done on a sample
closer to the bottom of the line.  In Figures
\ref{CaII.eps}-\ref{FeIII1_vel.eps} we show the time evolution of the
velocities for Ca~{\sc ii} H\&K, Si~{\sc ii}~$\lambda$6355, and
Fe~{\sc iii}~$\lambda\lambda$4404,5129 lines for SN~1999aa along with
several other supernovae from the literature.

Ca~{\sc ii} H\&K velocities for SN~1999aa, shown in
Fig. \ref{CaII.eps}, are in the range of those of normal SNe. This
feature shows an evident split in the minimum at day $-$3 and $-$1. In
these cases the measurements have been performed selecting a
wavelength range that covers the whole line profile from approximately
3600~\AA\ to 3850~\AA. This was chosen in order to obtain a value of
the velocity that could be compared with the other
supernovae. Measuring the two minima and deriving from those the
velocities of the two components of Ca~{\sc ii} used in the synthetic
spectrum would not be appropriate since the position of the troughs
are the result of the blending of Ca~{\sc ii}, Si~{\sc ii}, Co~{\sc
ii} and Ni~{\sc ii}.

Si~{\sc ii} velocities are shown in Fig.  \ref{siII635_vel.eps}.  If
the absorption near 6150~\AA\ of SN~1999aa was due only to Si~{\sc
ii}~$\lambda$6355, the velocities would have been monotonically
decreasing with time, as is seen for the normal SN~1994D, SN~1992A and
the sub-luminous SN~1999by and SN~1991bg. For SN~1999aa the first
point, 11 days before maximum light, has the lowest velocity. This is
consistent with another ion (probably C~{\sc ii} or H according to the
SYNOW fit, see sections \ref{synow-11_s} and \ref{carbonio}) also
being responsible for the absorption feature at this epoch. The
wavelength of the line minimum remains practically constant during the
first 20 days after maximum. This is usually interpreted as if the
element layer in the supernova atmosphere is confined to a region
above the photosphere, in this case around $10100$~km~s$^{-1}$ as
confirmed in the SYNOW synthetic spectra. The same trend is shown by
SN~2000cx and SN~1991T, with velocities around 12000~km~s$^{-1}$ and
9400~km~s$^{-1}$, respectively, as indicated by the dashed horizontal
lines in Fig.  \ref{siII635_vel.eps}.

The velocities measured for the two Fe~{\sc iii} features are shown in
Fig. \ref{FeIII_vel.eps} and \ref{FeIII1_vel.eps}. These lines are
characteristic of the pre-maximum spectra of the SN~1991T-like SNe~Ia
and disappear within the first week after maximum light. In both cases
SN~1999aa shows lower velocities compared to SN~1991T, SN 1997br and
SN~2000cx, although all have a similar slope. Fe~{\sc
iii}~$\lambda$4404 velocities (Fig. \ref{FeIII_vel.eps}), are smaller
by 2500~km~s$^{-1}$ with respect to SN~1991T, and 3500~km~s$^{-1}$
with respect to SN~2000cx.  Fe~{\sc iii}~$\lambda$5129 velocities
(Fig. \ref{FeIII1_vel.eps}) are still smaller, but much closer to the
velocities of the other peculiar supernovae. This could be due to a
higher contamination from Si~{\sc ii}~$\lambda$5051 or Fe~{\sc
ii}~$\lambda$5018.
 
Generally, the velocity ranges of the lines of SN~1999aa are 
consistent with those of normal supernovae. The major peculiarity is
the restricted atmosphere region where Si~{\sc ii} is present which makes
this ion appear clearly only one week before maximum light.

\section{Discussion}
\label{sec_summary}
The analysis carried out in section \ref{sec_comparison} indicates
that SN~1999aa was a peculiar object prior to maximum light,
developing toward normal-looking spectra around maximum. The SYNOW
synthetic spectra have shown the velocity ranges in which each
adopted ion appears, and offer a model of the structure of the
expanding atmosphere. In this section we summarize our findings for
different ions and their velocity distributions and discuss the
consequences for constraining the progenitor system and the explosion
models.

\subsection{Intermediate mass elements}
The velocity range in which IMEs are found constrains the atmospheric
region in which incomplete Si burning takes place, and thus the
hydrodynamics of the supernova explosion. We identify the
contributions of ions such as Si~{\sc ii}, S~{\sc ii}, Ca~{\sc ii},
Na~{\sc i} and Mg~{\sc ii} which are common to both normal and
SN~1991T-like supernovae. However, the optical depths and velocity
ranges found for these elements for SN~1999aa differ from those
typically used for Branch-normal SNe. The presence of unburned C and
IMEs (HV Ca~{\sc ii} and possible signs of C~{\sc ii}) above 14000
km~s$^{-1}$, as found in our synthetic spectra of SN~1999aa, suggests
that this supernova was not the result of a sub-Chandrasekhar mass
explosion, as the composition resulting from the modeling of such
explosions do not include any IME or C in the external layer of the
envelope (see e.g., \markcite{1995ApJ...452...62L}{Livne} \& {Arnett}
(1995) and \markcite{1994ApJ...423..371W}{Woosley} \& {Weaver}
(1994)).

The time evolution of the Doppler blue-shift of Si~{\sc ii},
Fig. \ref{siII635_vel.eps}, around and after B-band maximum shows that
this ion is confined within a narrow velocity range (v$_{\rm min}
\sim$10100~km~s$^{-1}$), not only for SN~1999aa but also in SN~2000cx
(v$_{\rm min} \sim$12000~km~s$^{-1}$) and SN~1991T (v$_{\rm min}
\sim$9400~km~s$^{-1}$). Our SYNOW models of SN~1999aa provide
independent evidence for a confined Si~{\sc ii} layer.  Velocities
between $10000$~km~s$^{-1}$ and $15000$~km~s$^{-1}$ were needed to
match the observations when the photospheric velocity dropped below
$9500$~km~s$^{-1}$, (see Tables \ref{table-11} to \ref{table+40}).
The same holds for SN~1997br \markcite{2002NewA....7..441H}({Hatano}
{et~al.} 2002) and for SN~1991T
\markcite{1999MNRAS.304...67F}({Fisher} {et~al.} 1999). For a normal
Type Ia supernova, such as SN~1994D, the Si-rich layer extends from
the photosphere out to v~$>$~25000~km~s$^{-1}$,
\markcite{1999ApJ...525..881H}({Hatano} {et~al.} 1999a). For
SN~1991T-like or SN~1999aa-like objects, the weakness of the IME's
absorption and the earlier than normal domination of the Fe-group
elements suggest that the region of incomplete Si-burning could be
confined to a small velocity window. Thus, a higher deflagration to
detonation transition density in DD models could explain some of the
spectral peculiarities.

In delayed detonation hydrodynamical models this would imply an
over-luminous object.  We do not have a sufficiently accurate absolute
magnitude measurement of SN~1999aa, but if the light curve-width
luminosity relation holds for this object, SN~1999aa should be more
luminous than normal SNe. Based on the measurement of the distance to
SN~1991T, \markcite{2001ApJ...551..973S}{Saha} {et~al.} (2001) and
\markcite{2001ApJ...547L.103G}{Gibson} \& {Stetson} (2001) showed that
that supernova is not necessarily overluminous. Thus, the possible
correlations between the absolute magnitude of SNe~Ia and the velocity
range in which IMEs are present requires further studies.

\subsection{Nickel and other doubly ionized elements}
The ionization level of each component of the atmosphere in type Ia
SNe is an indicator of the energy balance during the supernovae
explosion. Both thermal and non-thermal ionization are known to
be important in type Ia SNe (see
e.g. \markcite{1991ApJ...383..308L,1995ApJ...455L.147N,1996MNRAS.283..297B}{Lucy}
(1991); {Nugent} {et~al.} (1995); {Baron} {et~al.} (1996)).  

We have found evidence for Si~{\sc iii}, Fe~{\sc iii} and Ni~{\sc iii}
in the early spectra of SN~1999aa. As discussed in
\markcite{1995ApJ...455L.147N}{Nugent} {et~al.} (1995), doubly ionized
Fe and Si in the early spectra suggest higher temperature than in
Branch-normal SNe. The absorption features from these ions disappear
after day 5, as the supernova atmosphere cools.

The innermost layers of deflagration models, such as W7, are
considered to have gone through complete silicon burning, leaving only
a nuclear statistical equilibrium composition such as that shown in
\markcite{1999ApJS..121..233H}{Hatano} {et~al.} (1999b). The decay of the resulting $^{56}$Ni to
$^{56}$Co and then to stable $^{56}$Fe would imply the presence of
Co~{\sc iii} lines not found in the present analysis. As in
\markcite{2002NewA....7..441H}{Hatano} {et~al.} (2002), we suggest that the Ni~{\sc iii} lines in
our pre-maximum spectra are produced by $^{54}$Fe and $^{58}$Ni
synthesized during incomplete and complete silicon burning. If this is
the case, it is not necessary to produce a model that extends the
complete silicon burning into regions of intermediate mass elements or
above, as originally proposed for SN~1991T-like objects. Among the
current DD models, none is able to extend the presence of these ions
sufficiently far out (v$_{\rm max}=30000 $~km~s$^{-1}$, Table
\ref{table-11}) to match our observations.

Furthermore, the measurements of the Doppler shift of Fe~{\sc iii},
Fig. \ref{FeIII_vel.eps} and \ref{FeIII1_vel.eps}, show a lower value
than other SN~1991T-like objects. This suggests that Fe~{\sc iii}
extends further out in SN~1999aa than in spectroscopically normal
SNe~Ia, but has its maximal absorption at lower radii than 
other SN~1991T-like objects.

\markcite{1998ApJ...495..617H}{Hoeflich}, {Wheeler}, \& {Thielemann}
(1998) showed that a higher initial metallicity would increase the
abundance of $^{54}$Fe and $^{58}$Ni, and this could possibly explain
our observations.  UV photometry and spectroscopy would yield
important information to settle this issue and aid in the development
of new models that could better match the observations
\markcite{2000ApJ...530..966L}({Lentz} {et~al.} 2000).
\subsection{Carbon~{\sc ii} or Hydrogen?}
\label{carbonio}

The identification of H or C would impose constraints on either the
nature of the progenitor system or the hydrodynamics of the explosion
respectively. In section \ref{synow-11_s} we suggested the presence of
a HV C (C~{\sc ii}) component for the spectrum of SN~1999aa
at day $-$11. In what follows, we discuss possible alternative
explanations, assuming an additional ion is necessary to improve the
matching.

The evidence for an external layer of carbon in Type Ia supernovae has
been discussed in the case of SN~1991T by \markcite{1997ApJ...481L..89F}{Fisher} {et~al.} (1997)
and \markcite{1999MNRAS.304...67F}{Fisher} {et~al.} (1999), for SN~1990N by
\markcite{2001MNRAS.321..341M}{Mazzali} (2001), for SN~1994D by
\markcite{1999ApJ...525..881H}{Hatano} {et~al.} (1999a) and recently for SN~1998aq by
\markcite{2003AJ....126.1489B}{Branch} {et~al.} (2003), based mainly on the possible presence of
a C~{\sc ii} absorption on the red side of Si~{\sc
ii}~$\lambda$6355. For SN~1999aa, this ion helps in reproducing the
line shape and would explain the velocity time evolution of the
feature at $6150$~\AA , as shown in Figs.~\ref{synow-11} and
\ref{synow-11CII}, and Fig. \ref{siII635_vel.eps} (section
\ref{sec_velocities}).  However, due to the small optical depth for
C~{\sc ii} ($\tau=0.01$, table \ref{table-11}) the C~{\sc ii} 7234
line, which could confirm the presence of C, is too weak in the
synthetic spectrum to be positively identified in the spectrum
observed at day $-$11.

Our identification of an external layer of C in SN~1999aa remains
tentative (based only on the contribution at 6150~\AA), but seems
plausible given the work of other authors on early spectra of
SN~1991T-like and Branch-normal SNe.  The presence of C would point to
explosion hydrodynamics consistent with pure deflagration of a
Chandrasekhar mass WD (e.g. \markcite{1984ApJ...286..644N}{Nomoto},
{Thielemann}, \& {Yokoi} (1984)).

\begin{figure*}
\centering 
 \includegraphics[width=16cm]{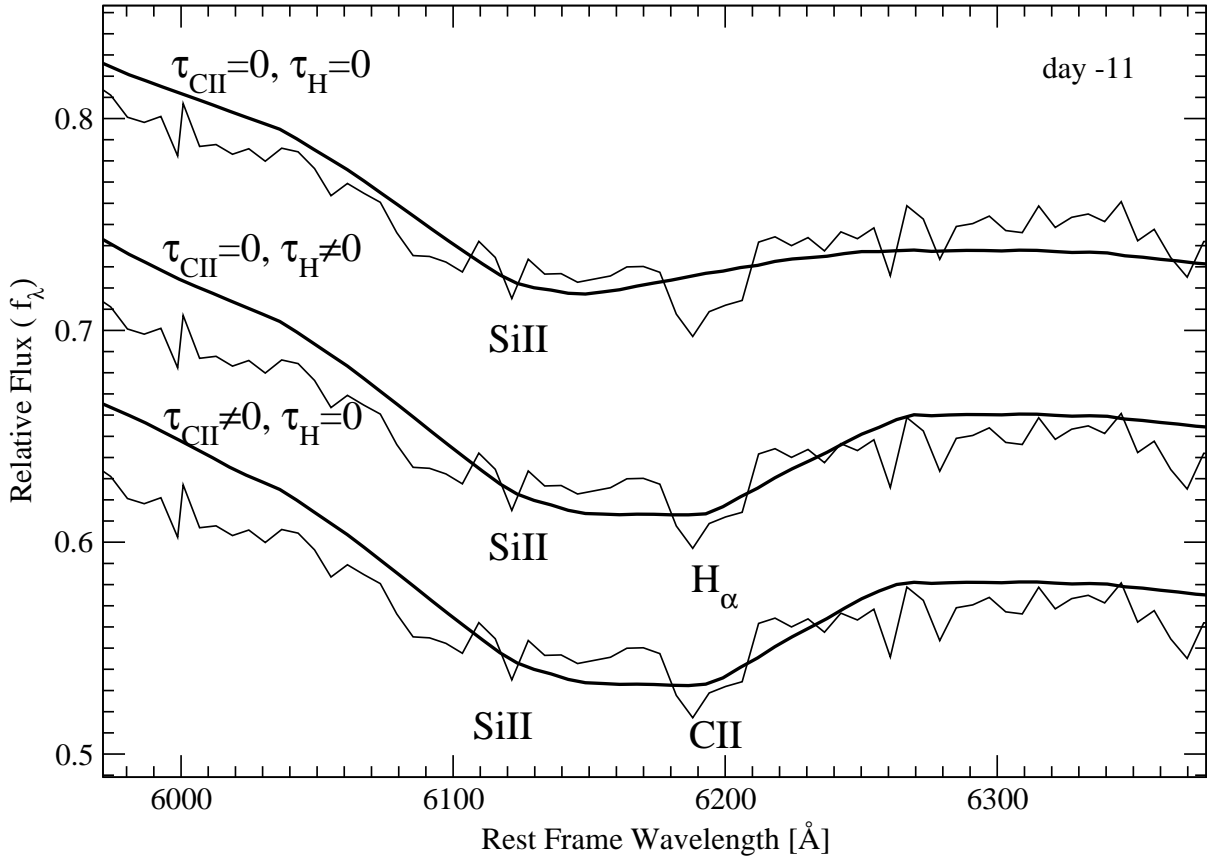}
  \caption{Synthetic spectra compared with SN~1999aa spectrum for day
  $-$11 around 6150~\AA. First model from the top: Solid lines,
  $\tau_{\rm H}=0$, $\tau_{\rm CII}=0$ and data. Second model from the
  top: Solid lines, $\tau_{\rm H}\neq0$, $\tau_{\rm CII}=0$ and
  data. Third model from the top: Solid lines, $\tau_{\rm
  CII}\neq0$,$\tau_{\rm H}=0$  and data.  The
  continuum level has been shifted to match the data. SYNOW parameters
  for C~{\sc ii} are presented in table \ref{table-11} and those for H
  are presented in the text. Ions responsible for features in the
  synthetic spectrum are labeled. }
  \label{CIIalter}
\end{figure*}

\markcite{2002ApJ...580..374L}{Lentz} {et~al.} (2002) discussed the possibility that the
absorption redward of Si~{\sc ii}~$\lambda$6355 could be attributed to
hydrogen mixed into the external layer from the WD companion. They
modeled the presence of solar composition material mixed in the
unburned C+O layer of W7 to study the effect of different mixing
depth. In some configurations they found that H$\alpha$ could be
easily mistaken for C~{\sc ii}~$\lambda$6580. We have explored this
possibility assuming the applicability of the resonance-scattering
approximation, as illustrated in Fig. \ref{CIIalter}.

The continua of the three models proposed have been shifted (adding an
arbitrary constant) to match the data.  The model on the top of
Fig. \ref{CIIalter} includes only Si~{\sc ii} and, as noted in section
\ref{synow-11_s}, the minimum velocity is set by the photosphere
velocity which is constrained by other ions to be v$_{\rm
phot}$=11$\times 10^3 $ km~s$^{-1}$ and cannot be changed without
adversely affecting the overall match.  The model in the middle
reproduces better the line profile thanks to the contribution of the
H$\alpha$ line.  The parameters used for H are: v$_{\rm
min}$=18$\times 10^3 $ km~s$^{-1}$, v$_{\rm max}$=30$\times 10^3
$km~s$^{-1}$, $\tau$=0.08, T$_{\rm exc}$=15$\times 10^3$K and
v$_e$=5$\times 10^3 $km~s$^{-1}$. The optical depth used does not form
any visible H$\beta$ line around 4500~\AA, which is in agreement with
our observations. Note, however, that for resonance scattering the net
emission phenomenon is neglected and this could change the relative
strength of H$\alpha$ and H$\beta$ (see
e.g. \markcite{2000ApJ...545..444B,2003astro.ph..2260T}{Baron} {et~al.} (2000); {Thomas} {et~al.} (2003)).  The
reproduction of the line profile at 6150~\AA\ is indeed as good for
H$\alpha$ as for C~{\sc ii} (bottom of Fig. \ref{CIIalter}). The
identification of a given ion in a supernova atmosphere based on the
evidence of a single line is not definitive, therefore we consider the
contribution from either of the two ions equally plausible.

The presence of hydrogen in the external layers of the exploding WD
would indicate that a single degenerate progenitor system scenario is
responsible for at least some of the observed SNe~Ia.

\subsection{Carbon~{\sc iii} or Hydrogen?}
In section \ref{synow-11_s} we included C~{\sc iii}~$\lambda$4648.8 at
photospheric velocities to match the line at 4500~\AA.  The presence
of unburned material deep into the IME atmospheric region is seen in
3D deflagration models, where convective flows are shown to
continuously refuel the inner layers with material that has original
composition \markcite{2003Sci...299...77G}({Gamezo} {et~al.} 2003). Finding unburned material
deep into the atmosphere is then theoretically possible.

The first attempt to identify this line as C~{\sc iii} was performed
by \markcite{2002NewA....7..441H}{Hatano} {et~al.} (2002) using the early spectra of SN~1997br, but they
could not match the central wavelength of the line.  They
alternatively mentioned the possibility that He~{\sc ii} could be
responsible for the absorption. However, because of the high
ionization energy of helium this could only be the case if there was a
large degree of non-thermal ionization from the decay products of
radioactive elements.

\begin{figure*}
\centering 
 \includegraphics[width=16cm]{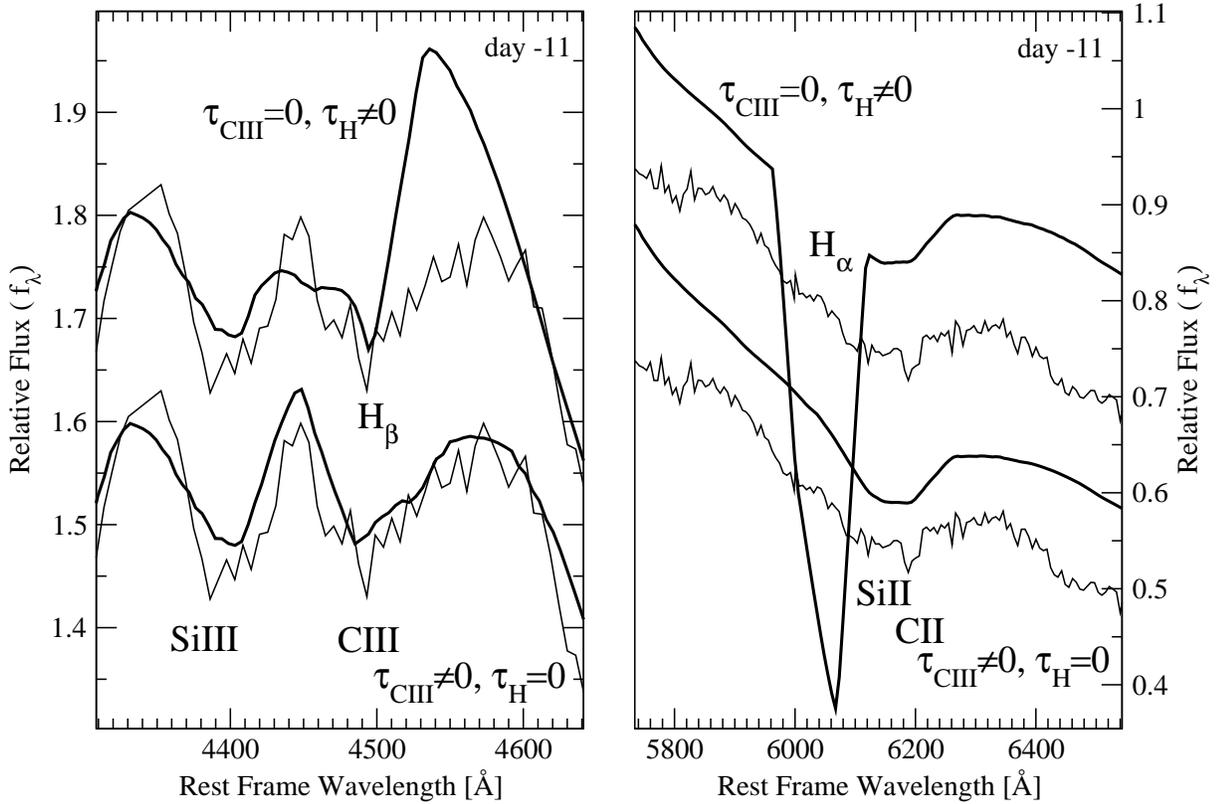}
  \caption{Synthetic spectra compared with SN~1999aa spectrum for day
  $-$11 around 4500~\AA\ (left panel) and 6150~\AA\ (right
  panel). First model from the top: Solid lines, $\tau_{\rm H}\neq0$,
  $\tau_{\rm C III}=0$ and data. Second model from the top: Solid lines,
  $\tau_{\rm H}=0$, $\tau_{\rm C III}\neq0$ (as second model in
  Fig. \ref{synow-11CIII}) and data.  In the resonance scattering
  approximation the optical depth necessary to reproduce the observed
  absorption at 4500~\AA\ with H$\beta$ produces a too strong
  H$\alpha$ that could be weaken considering the net emission effect.
  SYNOW parameters for C~{\sc iii} are presented in table
  \ref{table-11} and those for H are presented in the text. Ions
  responsible for features in the synthetic spectrum are labeled.}
  \label{CIIIalter}
\end{figure*}

\markcite{2003astro.ph..2260T}{Thomas} {et~al.} (2003) tentatively suggested the possibility that
the same line in SN~2000cx could be matched by a HV clump of H. In our
spectrum at day 11 prior to maximum this would not be possible because
the optical depth necessary to match the line at 4500~\AA\ with
H$\beta$ would produce H$\alpha$ in the Si~{\sc ii}~$\lambda$6355~\AA\
region, which is ruled out by the data as shown in Fig. \ref{CIIIalter}
(SYNOW parameters for H: v$_{\rm min}$=24.5$\times 10^3 $km~s$^{-1}$,
v$_{\rm max}$=30$\times 10^3 $km~s$^{-1}$, $\tau$=2.5, T$_{\rm
exc}$=15$\times 10^3$K and v$_e$=3$\times 10^3
$km~s$^{-1}$). However, note that considering the net emission effect
would weaken the H$\alpha$ absorption feature. Furthermore, as pointed
out in \markcite{2003astro.ph..2260T}{Thomas} {et~al.} (2003), because of the lack of a strong
emission in the 6150~\AA\ region, the identification of the line at
4500~\AA\ as H$\beta$ would be contingent upon a clumpy geometry and
thus  requires a 3D simulation to be fully explored.

Because of the good line profile fit, C~{\sc iii} remains our best
candidate for the identification of the line at 4500~\AA\ in the early
spectra of SN~1999aa. Such a possibility would strongly favor current
3D deflagration models of C+O WD since they produce this ion at all
velocities,
\markcite{2003Sci...299...77G,2003ApJ...588L..29B,Khokhlov:2000gj}({Gamezo} {et~al.} 2003; {Baron} {et~al.} 2003; Khokhlov 2000).
Further confidence in this identification will require detailed
radiative transfer modeling.
\subsection{High velocity Ca~{\sc ii}}
The identification of high velocity burned material in supernova
spectra can place constraints on the hydrodynamics of the
explosion. The presence of a HV component of Ca~{\sc ii} is
strongly supported by the blue components of Ca~{\sc ii}~IR triplet
(Fig. \ref{synow-1}). In the synthetic spectrum, the strength of the
bluest of the two HV absorptions features is almost twice
that of the red one. This is not the case in the observed feature and
could be a sign of asymmetry in the high velocity ejecta. To reproduce
the correct flux level, a three dimensional study of the geometry of
the atmosphere (such as that computed for SN~2001el in
\markcite{2003ApJ...593..788K}{Kasen} {et~al.} (2003)) would be necessary. An optically thick
HV clump viewed from a line of sight such that it just
barely covers the photosphere can produce a weak IR triplet feature
with two minima of equal depth \markcite{2003ApJ...593..788K}({Kasen} {et~al.} 2003) such as
seen in SN~1999aa.

Evidence for a HV component of Ca~{\sc ii} also comes from the
parameters in our synthetic spectra needed to model observed Ca~{\sc
ii} H\&K . To reproduce completely the broad profile of this line at
day $-$11 and the blue minimum at $-$1 days the HV component is
needed. When the temperature of the atmosphere drops, and the
photosphere recedes into the inner layers, low ionization Fe-peak
group lines become more important in the Ca H\&K region. In this case,
reproducing the line profile accurately becomes difficult. However, we
could not match the Ca~H\&K absorption shown at day $-$11 as well as
at day $-$1 by any means other than a HV Ca~{\sc ii} component. The
evolution of this feature shows that the red component becomes
stronger, and that by three weeks after maximum it dominates the
entire feature. This would be consistent with a decreasing optical
depth of the HV Ca H\&K as the atmosphere expands. This is consistent
with our interpretation, namely that the HV Ca~{\sc ii} component
could be formed in a clump.

Recent spectropolarimetry measurements of type Ia supernovae
\markcite{2001ApJ...556..302H,2003ApJ...591.1110W,2003ApJ...593..788K}({Howell} {et~al.} 2001; {Wang} {et~al.} 2003; {Kasen} {et~al.} 2003)
have shown that, when present, the HV component of the Ca~{\sc
ii}~IR triplet is usually polarized. The study of the polarization
parameters can yield important information on the geometry of the
explosion and thus can help in the development of 3D models.

\section{Conclusions}
\label{sec_conclusion}
We have presented new high signal to noise optical spectroscopic data
of SN~1999aa with good temporal coverage between $-$11 and +58 days
with respect to B-band maximum light.  The overall evolution of the
spectral features suggests an object with characteristics common to
both SN~1991T-like and Branch-normal supernovae.

\begin{figure*}
\centering 
 \includegraphics[width=16cm]{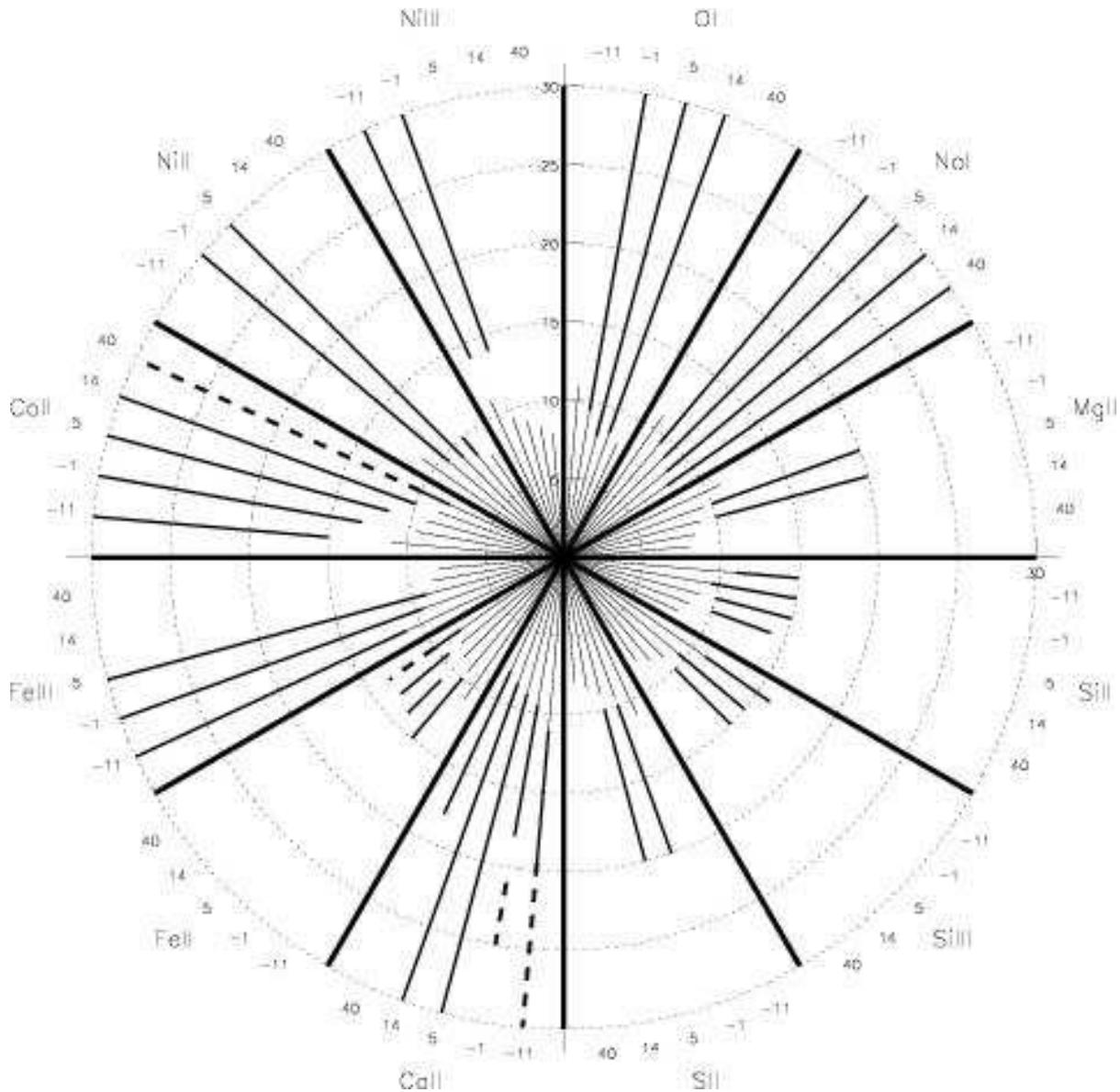}
  \caption{Schematic representation of the composition of SN~1999aa in
  velocity space as inferred from the SYNOW synthetic spectra of day
  $-$11, $-$1, +5, +14 and +40. Velocities increase radially in units
  of 10$^3$ km~s$^{-1}$. The velocity ranges (v$_{min}$-v$_{max}$) for
  12 ion species used in the synthetic spectra are marked with solid
  lines inside the circle section labeled with the ion name. High
  velocity components are marked with dashed lines. In each ion
  section the five epochs analyzed are shown. For each epoch the
  extent of the photosphere is marketed with a thin solid line. The
  contributions of C~{\sc ii}, C~{\sc iii}, [O~{\sc ii}] and H are not
  shown. For details see section \ref{sec_comparison}.}
  \label{onion}
\end{figure*}

By means of SYNOW synthetic spectra we have attempted to identify the
absorption features in the spectra at days $-$11, $-$1, +5, +14 and
+40 with respect to B-band maximum light. Highlights of this
modeling include the presence of
\begin{itemize}
\item C~{\sc iii} at the photospheric velocity, 
\item Possible C~{\sc ii} or H at high velocity
\item Ca~{\sc ii}~IR triplet at high velocity, 
\item doubly ionized Si, Ni and Fe, 
\item confined IMEs and Fe~{\sc ii}, 
\item probable iron peak core at 10000 km~s$^{-1}$. 
\end{itemize}
A schematic view of the resulting composition (in the velocity regime)
is presented in Fig. \ref{onion} .

The line identification for the earliest spectrum shows the presence
of doubly ionized Si, Fe and Ni suggesting that the temperature in the
outermost layer of SN~1999aa is higher than in normal SNe. After day 5,
doubly ionized elements no longer form visible absorption features,
highlighting the cooling of the supernova atmosphere.  The presence of
IME above 14000 km~s$^{-1}$ excludes the possibility that SN~1999aa
was the result of a sub-Chandrasekhar mass explosion.

The broad absorption feature around 6150~\AA\ in the spectrum at day
$-$11 can be matched comparably well by a weak component of Si~{\sc
ii}~$\lambda$6355 plus HV C~{\sc ii}~$\lambda$6580 or
H$\alpha$. The contribution of either C~{\sc ii} or H$\alpha$ helps in
reproducing the velocity time evolution of the absorption feature at
6150~\AA\ that otherwise (i.e. considering only Si~{\sc
ii}~$\lambda$6355) would remain unexplained. Definitive identification
is not possible due to the lack of other absorption features produced
either by C~{\sc ii} or H at the input optical depths. However, both
the possibilities impose constraints either on the nature of the
progenitor system or the hydrodynamics of the explosion. A hydrogen
component would imply that SN~1999aa is the result of a single
degenerate WD explosion; a complete 3D calculation would be necessary
to model it. Alternatively, a C~{\sc ii} component can be reproduced
by a pure 1D deflagration model, but would not be consistent with
delayed-detonation models.

C~{\sc iii}~$\lambda$4648.8 at photosphere velocity was able to
reproduce the absorption around 4500~\AA, and no good alternatives
were found for that feature. If this identification is confirmed by
detailed radiative transfer models, the use of three dimensional
simulations of the explosion would be necessary to describe the
presence of unburned material down to the inner layers of the WD
atmosphere.

At day $-$11 and at day $-$1, the profile of the absorption at
3800~\AA\ requires a HV component of Ca to be reproduced in
addition to the photosphere component. The HV component at
day $-$1 also reproduces the observed minima on the blue side of the
usual Ca~IR triplet. For an accurate description of this feature,
multi-dimensional simulations would be required.

The spectral evolution of SN~1999aa in the photospheric phase shows
that Fe~{\sc ii} is confined below 15000 km~s$^{-1}$. IMEs populate a
narrow velocity window above 10000 km~s$^{-1}$.  Similar evidence is
found in other well known supernovae (SN~1991T, SN~1997br and
SN~2000cx) by studying the time evolution of the expansion velocity as
computed from fits to the minima of Si~{\sc ii}~$\lambda$6355. The
comparison with other SN~1991T-like objects suggests that the
transition between IME to iron-peak dominant composition can occur at
slightly different phases. Tuning of the deflagration to detonation
transition density might reproduce this sequence.

The analysis of the spectrum of day +40 indicates that the iron-peak core
limit should be set to 10000 km~s$^{-1}$, similar to that of SN~1991T
\markcite{1999MNRAS.304...67F}({Fisher} {et~al.} 1999).

The origin of the differences between normal supernovae and
SN~1991T-like or SN~1999aa-like objects depends on several factors.  Our
analysis of the optical spectra of SN~1999aa reveals that, among the
present explosion models, none is able to reproduce each one of our
findings. Higher temperature could account for some of the
peculiarities \markcite{1995ApJ...455L.147N}({Nugent} {et~al.} 1995), but the evidence for high
velocity components and unburned material at all velocities probably
requires the development of full NLTE 3D explosion models. SN~1999aa
was spectroscopically less extreme than other genuine SN~1991T-like
SNe, suggesting that perhaps a single model could eventually explain
both normals and SN~1991T-like SNe as a continuous sequence.

\acknowledgements We would like to thank Rollin Thomas for helpful
comments and David Branch, Adam Fisher and Rollin Thomas for providing
the SYNOW code. We are grateful to the referee Mario Hamuy for useful
suggestions.  The research presented in this article made use of the
SUSPECT\footnote{ http://www.nhn.ou.edu/$\sim$suspect} Online
Supernova Spectrum Archive, and the atomic line list of
\markcite{1993KurCD...1.....K}{Kurucz} (1993).  This work is based on
observations made with: the Nordic Optical Telescope, operated on the
island of La Palma jointly by Denmark, Finland, Iceland, Norway, and
Sweden, in the Spanish Observatorio del Roque de los Muchachos of the
Instituto de Astrofisica de Canarias; the Apache Point Observatory
3.5-meter telescope, which is owned and operated by the Astrophysical
Research Consortium; the Lick Observatory Shane 3.0-m Telescope; the
Cerro Tololo Inter-American Observatory 4-m Blanco Telescope; the MDM
Observatory 2.4-m Hiltner Telescope.  We thank the telescope
allocation committees and the observatory staffs for their support for
this extensive observing campaign.  This work was supported in part by
"The Royal Swedish Academy of Sciences".  This work was supported in
part by the Director, Office of Science, Office of High Energy and
Nuclear Physics, of the U.S. Department of Energy under Contract
No. DE-AC03-76SF000098.  A. Mour\~ao acknowledges financial support
from Funda\c{c}\~ao para a Ci\^encia e Tecnologia (FCT), Portugal,
through project PESO/P/PRO/15139/99; S. Fabbro thanks the fellowship
grant provided by FCT through project POCTI/FNU/43749/2001.  Ariel
Goobar is a ``Royal Swedish Academy Research Fellow'' supported by a
grant from the Knut and Alice Wallenberg Foundation.

\clearpage 

\clearpage 
\clearpage 
\clearpage 
\clearpage 
\clearpage 
\clearpage 
\clearpage 
\clearpage 
\clearpage 
\clearpage 
\begin{figure*}
\centering 
 \includegraphics[width=16cm]{garavini.fig11.eps}
  \caption{Synthetic spectrum compared with SN~1999aa spectrum for day
  +5. SYNOW parameters used are presented in Table \ref{table+5}.
  Above 6200~\AA\ the black body assumption fails in reproducing the
  correct flux values but it does not affect the identification of the
  lines. Ions responsible for features in the
  synthetic spectrum are labeled.}
  \label{synow+5}
\end{figure*} 
\clearpage 
\clearpage 
\begin{figure*}
\centering
  \includegraphics[width=16cm] {garavini.fig13.eps}
  \caption{Synthetic spectrum compared with SN~1999aa spectrum for day
  +14. SYNOW parameters used are presented in Table \ref{table+14}.
  Above 6200~\AA\ the black body assumption fails in reproducing the
  correct flux values but it does not affect the identification of the
  lines. Ions responsible for features in the
  synthetic spectrum are labeled.}
  \label{synow+14}
\end{figure*} 
\clearpage 
\clearpage 
\clearpage 
\clearpage 
\clearpage 
\clearpage 
\clearpage 
\clearpage 
\clearpage 
\clearpage 
\clearpage 
\begin{table*}
\begin{center}
\caption{Data set specifications.\label{tab_data}}
\begin{tabular}{lrllcllr}
\tableline\tableline
JD & Epoch$\tablenotemark{f}$& Telescope &Instrument& $\lambda$ Range$\tablenotemark{f,g}$&$\langle
    \Delta\lambda\rangle$$\tablenotemark{a,f}$&$\langle S/N \rangle$$\tablenotemark{b}$& Comments\\ 
$-$2400000   & ref $B_{\rm max}$ & &  &[\AA] &[\AA]&\\
\tableline
51223.38 &  $-$11 & NOT 2.6m    & ALFOSC &3503-9295 & 6.1 & 52&5998 $\tablenotemark{c}$\\
51227.84 &  $-$7 &  APO 3.5m    & DIS   &3686-10268& 6.8 &  75&5490 $\tablenotemark{c}$\\
51231.35 &  $-$3  & NOT 2.6m    & ALFOSC &3508-9692 & 6.1 & 41&6097 $\tablenotemark{c}$\\
51234.71 &  $-$1  & Lick 3m & KAST &  3478-9975 & 3.1 &  43&5430 $\tablenotemark{c}$\\
51239.68 &  +5  & MDM 2.4m &MARK III &3872-8807& 5.3 &  45&$\tablenotemark{e}$\\
51239.68 &  +6  & MDM 2.4m &MARK III &3871-8808& 5.3 &  61&$\tablenotemark{e}$\\
51247.63 &  +14 & APO 3.5m & DIS& 3628-10333& 6.8 &  86&5887 $\tablenotemark{c}$\\
51253.37 &  +19 & NOT 2.6m    & ALFOSC &3375-9994& 6.0 &  63&6057 $\tablenotemark{c}$\\
51258.51 &  +25 & CTIO 4m  & RCSP&3320-9208 & 2.0 &  40&$\tablenotemark{d}$\\
51261.54 &  +28 & CTIO 4m  &RCSP &3374-9232& 2.1 & 42&$\tablenotemark{d}$\\
51266.53 &  +33 & CTIO 4m  &RCSP &3263-9199& 2.1 &  42&$\tablenotemark{d}$\\
51273.62 &  +40 & APO 3.5m & DIS  &3641-10115& 6.7 &  42&5877 $\tablenotemark{c}$\\
51282.67 &  +47 & Lick 3m  & KAST&3512-9940& 3.1 &  18&5462 $\tablenotemark{c}$\\
51286.63 &  +51 & APO 3.5m & DIS&3621-10166& 6.7 &  68&5874 $\tablenotemark{c}$\\
51293.71 &  +58 & Lick 3m  & KAST&3705-7963& 2.1 & 11&5390 $\tablenotemark{c}$\\
\tableline
\end{tabular}
\tablenotetext{a}{Average wavelength-bin size.}
\tablenotetext{b}{Average signal-to-noise ratio per wavelength bin.}
\tablenotetext{c}{Beginning of red channel, [\AA].}
\tablenotetext{d}{Negligible 2$^{nd}$ order contamination.}
\tablenotetext{e}{Possible 2$^{nd}$ order contamination above 7500~\AA.}
\tablenotetext{f}{In Rest Frame.}
\tablenotetext{g}{Where S$/$N $\ge$ 5.}
\end{center}
\end{table*}
\clearpage 
\begin{table*}
\begin{center}
\caption{SYNOW parameters for day $-$11. The fit is shown in Fig. \ref{synow-11}. ${\rm v}_{\rm
phot}$=11000~km~s$^{-1}$, $T_{\rm bb}=13700$~K. (HV) next to
Ca~{\sc ii} stands for high velocity component. \label{table-11}}
\begin{tabular}{llllll}
\tableline\tableline
Ion  & $\tau$ & ${\rm v}_{\rm min}$&${\rm v}_{\rm max}$&$T_{\rm exc}$ &${\rm v}_{\rm e}$\\
  &&$10^3$kms$^{-1}$&$10^3$km s$^{-1}$&\small$10^{3}$K &$10^3$km s$^{-1}$\\
\tableline
Ca~{\sc ii}&1.15&---&19.5&15&5\\
Ca~{\sc ii}~({\sc hv})&1.45&19.5&30&15&10\\
Si~{\sc ii}&0.1&---&15&15&5\\
Si~{\sc iii}&0.3&---&16&15&5\\
Co~{\sc ii}&0.25&14&30&10&5\\
Fe~{\sc iii}&0.63&---&30&10&5\\
Ni~{\sc iii}&2&14&30&10&5\\
C~{\sc ii}&0.01&19&30&15&5\\
C~{\sc iii}&0.5&---&13.5&15&5\\
\tableline
\end{tabular}
\end{center}
\end{table*}
\clearpage 
\begin{table*}
\begin{center}
\caption{SYNOW parameters for day $-$1. The fit is shown in Fig. \ref{synow-1}. ${\rm v}_{\rm phot}$=9500~km~s$^{-1}$, $T_{\rm bb}=13500$~K. (HV) next to
Ca~{\sc ii} stands for high velocity component.\label{table-1}}

\begin{tabular}{llllll}
\tableline\tableline
Ion & $\tau$ & ${\rm v}_{\rm min}$&${\rm v}_{\rm max}$&$T_{\rm exc}$ &${\rm v}_{\rm e}$\\ 
 & &$10^3$kms$^{-1}$ & $10^3$km s$^{-1}$ &$10^{3}$ K & $10^3$km s$^{-1}$ \\
\tableline
Ca~{\sc ii}&3&---&18&8&5\\ 
Ca~{\sc ii}~({\sc hv})&2.5&22&25&10&5\\ 
O {\sc i}&0.2&---&30&8&5\\
Si~{\sc ii}&0.9&---&15&8&5\\ 
Si~{\sc iii}&0.25&---&15&8&5\\
S~{\sc ii}&0.65&10&20&8&5\\
Mg~{\sc ii}&0.1&10&20&8&5\\
Fe~{\sc ii}&1&10&15&8&5\\
Co~{\sc ii}&0.3&13&30&8&5\\ 
Na {\sc i}&0.25&---&30&8&5\\ 
Fe~{\sc iii}&0.8&---&30&8&3\\
Ni~{\sc ii}&0.1&---&30&8&5\\ 
\tableline
\end{tabular}
\end{center}
\end{table*}
\clearpage 
\begin{table*}
\begin{center}
\caption{SYNOW parameters for day +5. The fit is shown in Fig. \ref{synow+5}. ${\rm v}_{\rm phot}$=9000~km~s$^{-1}$, $T_{\rm bb}=12000$~K.\label{table+5}}

\begin{tabular}{llllll}
\tableline\tableline
Ion  & $\tau$ & ${\rm v}_{\rm min}$&${\rm v}_{\rm max}$&$T_{\rm exc}$ &${\rm v}_{\rm e}$\\
   & &$10^3$kms$^{-1}$ &$10^3$km s$^{-1}$ &$10^{3}$K & $10^3$km s$^{-1}$ \\
\tableline
Ca~{\sc ii}&20&---&30&5&3\\
O~{\sc i}&0.1&---&30&5&3\\
Si~{\sc ii}&1.5&10&15&5&3\\
Si~{\sc iii}&0.05&10&15&5&3\\
S~{\sc ii}&0.45&10&20&5&3\\
Mg~{\sc ii}&0.6&10&20&5&3\\
Fe~{\sc ii}&1.3&11&14&5&3\\
Co~{\sc ii}&0.8&11.5&30&5&3\\
Na {\sc i}&0.4&---&30&5&5\\
Fe~{\sc iii}&0.09&---&30&5&3\\
Ni~{\sc ii}&0.2&---&30&5&3\\
\tableline
\end{tabular}
\end{center}
\end{table*}
\clearpage 
\begin{table*}
\begin{center}
\caption{SYNOW parameters for day +14. The fit is shown in Fig. \ref{synow+14}. ${\rm v}_{\rm phot}$=8500~km~s$^{-1}$, $T_{\rm bb}=10500$~K.\label{table+14}}

\begin{tabular}{llllll}
\tableline\tableline
Ion  & $\tau$ & ${\rm v}_{\rm min}$&${\rm v}_{\rm max}$&$T_{\rm exc}$ &${\rm v}_{\rm e}$\\
 & &$10^3$kms$^{-1}$& $10^3$km s$^{-1}$&$10^{3}$K  & $10^3$km s$^{-1}$\\
\tableline
Ca~{\sc ii}&20&---&30&5&3\\
O~{\sc i}&0.1&---&30&5&3\\
Si~{\sc ii}&1.5&10&14&5&3\\
Fe~{\sc ii}&4&9.5&13.5&5&3\\
Co~{\sc ii}&1.7&10&30&5&3\\
Na~{\sc i}&1.2&---&30&5&5.5\\
Ni~{\sc ii}&5&---&10&5&3\\
\tableline
\end{tabular}
\end{center}
\end{table*}
\clearpage
\begin{table*}
\begin{center}
\caption{SYNOW parameters for day +40. The fit is shown in
Fig. \ref{synow-11}. ${\rm v}_{\rm phot}$=8000~km~s$^{-1}$, $T_{\rm
bb}=8000$~K.\label{table+40}}

\begin{tabular}{llllll}
\tableline\tableline
Ion  & $\tau$ & ${\rm v}_{\rm min}$&${\rm v}_{\rm max}$&$T_{\rm exc}$ &${\rm v}_{\rm e}$\\
 & &$10^3$kms$^{-1}$ & $10^3$km s$^{-1}$ &$10^{3}$K & $10^3$km s$^{-1}$ \\
\tableline
Ca~{\sc ii}&300&---&18&8&3\\
O~{\sc ii}&1&---&30&10&10\\
Fe~{\sc ii}&160&---&10&5&3\\
Fe~{\sc ii}&0.2&10&13.5&5&3\\
Co~{\sc ii}&60&---&10&5&3\\
Co~{\sc ii}&0.2&10&30&5&3\\
Na~{\sc i}&0.9&---&30&5&5\\
\tableline
\end{tabular}
\end{center}
\end{table*}
\end{document}